\documentclass[]{aa520}

\usepackage{amsmath,amssymb,graphicx,mathptmx,natbib}

\authorrunning{Puchwein and Bartelmann}
\titlerunning{Three-dimensional reconstruction of the intra-cluster medium}

\begin{document}

\title{Three-dimensional reconstruction of the intra-cluster medium}
\author{Ewald Puchwein and Matthias Bartelmann
    \institute{Zentrum f\"ur Astronomie der Universit\"at
    Heidelberg, ITA, Albert-\"Uberle-Str.~2, 69120 Heidelberg,
    Germany}}

\date{\emph{Astronomy \& Astrophysics, submitted}}

\abstract{We propose and test a new method based on Richardson-Lucy
  deconvolution to reconstruct three-dimensional gas density and temperature
  distributions in galaxy clusters from combined X-ray and thermal
  Sunyaev-Zel'dovich observations. Clusters are assumed to be axially
  symmetric and arbitrarily inclined with respect to the line-of-sight. No equilibrium assumption other than local thermal equilibrium is needed.
  We test the algorithm with synthetic observations of analytically
  modeled and numerically simulated galaxy clusters and discuss the
  quality of the density and temperature reconstructions in idealised
  situations and in presence of observational noise, deviations from
  axial symmetry and cluster substructure. We find that analytic and
  numerical gas density and temperature distributions can be accurately
  reconstructed in three dimensions, even if observational noise is
  present. We also discuss methods for determining the inclination
  angle from data and show that it can be constrained using X-ray
  temperature maps. For a realistic cluster and including observational noise the three-dimensional reconstructions reach a level of accuracy of about 15$\%$.
}
\maketitle

\section{Introduction}

In hierarchical models of structure formation, galaxy clusters are not
only the most massive gravitationally bound objects in the Universe,
but also the most recently forming. Numerous examples show that they
are typically irregularly shaped and occasionally undergoing violent
merger events. Cluster-sized dark-matter halos in simulations can
often be well described as triaxial ellipsoids, but not as spheres
\citep{JI02.1}.

At the same time, observations of galaxy clusters are often
interpreted based on spherically-symmetric models in hydrostatic
equilibrium. The beta model \citep{CA76.1} is still routinely being used for analyses
of the X-ray emission and also of the amplitude of the thermal
Sunyaev-Zel'dovich effect. Given the rapidly improving quality and
diversity of cluster data, it appears timely to search for an
algorithm which avoids the assumption of spherical symmetry and allows
the joint analysis of different types of cluster data.

Several such algorithms have been proposed. \cite{ZA98.1} suggested to
base the reconstruction of axisymmetric, three-dimensional
gravitational cluster potentials on the Fourier slice theorem,
extrapolating Fourier modes into the ``cone of ignorance''. They
applied their technique to simulated data and showed that it performs
well \citep{ZA01.1}. \cite{DO01.3} followed a perturbative approach,
and \cite{LE04.1} proposed to adapt parameters of triaxial halo
models, all by combining different data sets such as X-ray, (thermal)
Sunyaev-Zel'dovich (SZ) and gravitational-lensing maps. A similar method was applied to data by \cite{DE05.1}. An alternative
approach based on the iterative Richardson-Lucy deconvolution was
suggested by \cite{RE00.1} and \cite{RE01.1}. It aims at the gravitational potential, assumes only axial
symmetry of the main cluster body, avoids extrapolations in Fourier
space, and can easily be extended to include additional data sets.

In this paper, we develop the latter algorithm further. However, aiming at the
potential would require us to assume a relation between the gas
distribution and the gravitational field, which would be most conveniently
given by hydrostatic equilibrium. But even ignoring this common
equilibrium assumption, it should be possible to reconstruct the
three-dimensional distributions of intra-cluster gas density and
temperature by a joint analysis of X-ray and thermal SZ data.

We demonstrate here that this is indeed possible under the one
simplifying assumption that the underlying three-dimensional
distributions be axially (not spherically!) symmetric. The inclination of the symmetry axis can be arbitrary. We introduce the algorithm in Sect.~2 and apply it to the idealised case of an
analytically modeled, ellipsoidal cluster without substructure in
Sect.~3. Results obtained first without, then with observational noise are
highly promising: both the three-dimensional density and temperature
distributions are accurately reproduced. Noise suggests smoothing, and
we describe a suitable smoothing algorithm.

We study the less-ideal case of a numerically-simulated galaxy cluster
in Sect.~4. Here, axial symmetry is typically violated by the main
cluster body, and substructures are present which further perturb the
symmetry. Yet, faithful reconstructions are possible even in presence
of realistic noise.

Section~5 finally describes how inclination angles can be constrained
using temperature maps, and Sect.~6 summarises and discusses our
results.

\section{The deprojection algorithm}

\subsection{Deprojection of axisymmetric quantities using
  Richardson-Lucy deconvolution}
\label{sec:depro_quantity}

\begin{figure}[ht]
\scalebox{0.7}
{
\begin{picture}(0,0)
  \includegraphics{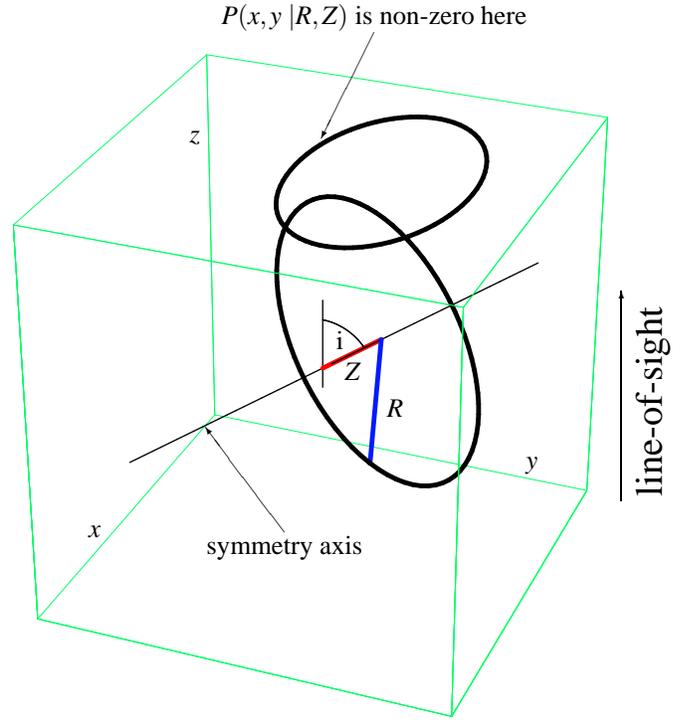}
\end{picture}
\setlength{\unitlength}{1mm}
\begin{picture}(140,140)(0,0)
  \put(60.5,71){\makebox(0,0){\strut{}\Large i}}
  \put(63,65){\makebox(0,0){\strut{}\Large $Z$}}
  \put(71,58){\makebox(0,0){\strut{}\Large $R$}}
  \put(14,35){\makebox(0,0){\strut{}\Large $x$}}
  \put(97,48){\makebox(0,0){\strut{}\Large $y$}}
  \put(33,110){\makebox(0,0){\strut{}\Large $z$}}
  \put (114,41){\vector(0,1){40}}
  \put(120,60){\rotatebox{90}{\makebox(0,0){\strut{}\huge line-of-sight}}}
  \put(67,130){\vector(-1,-2){10}}
  \put(67,133){\makebox(0,0){\strut{}\Large $P(x,y\;|R,Z)$ is non-zero here}}
  \put(50,35){\vector(-3,4){15}}
  \put(50,32){\makebox(0,0){\strut{}\Large symmetry axis}}
\end{picture}
}
\caption{Projection of an axisymmetric distribution of a physical
  quantity. The ellipse at the top marks the region where the kernel
  function corresponding to the projection along the line-of-sight is
  non-zero for fixed $R$ and $Z$.}
\label{fig:ellipse}
\end{figure}

As \cite{BI90.1} pointed out, Richardson-Lucy deconvolution
\citep{LU74.1,LU94.1} can be used to reconstruct an inclined
axisymmetric three-dimensional distribution of some physical quantity
$\phi$ from a two-dimensional map $\psi$ of its projection along the
line-of-sight. In astrophysical applications, $\psi$ will be data obtained from observations, for example the X-ray flux, the
lensing potential, or the Sunyaev-Zel'dovich decrement of an
approximately axisymmetric galaxy cluster. Because of the assumed
symmetry, $\phi$ can be written as a function of only two cylindrical
coordinates $R$ and $Z$, where we choose the symmetry axis as the
$Z$-axis (see Fig.~\ref{fig:ellipse}). Then, $R$ is the
distance from the symmetry axis. The projection along the
line-of-sight can be understood as a convolution of $\phi(R,Z)$ with a
kernel function $P(x,y|R,Z)$,
\begin{equation}
  \psi(x,y) \equiv \!\! \int_{-\infty}^{\infty} \!\!\!\!\!\! dz \; \phi(x,y,z)
            = \!\! \int_0^\infty \!\!\!\!\!\! \pi dR^2 \!\!
              \int_{-\infty}^{\infty} \!\!\!\!\!\! dZ \; \phi(R,Z) \; P(x,y|R,Z).
\label{eq:projection}
\end{equation}
The kernel function for a given pair $(R,Z)$ is non-zero only on the
ellipse obtained by projecting the ring onto the sky which is
defined by $R$ and $Z$ (see
Fig.~\ref{fig:ellipse}). It is derived in Appendix A of \cite{BI90.1}
and reads
\begin{equation}
  P(x,y|R,Z) = \frac{\delta[(\frac{y}{\cos i}-Z \tan i)^2-(R^2-x^2)]}{\pi\cos i},
\label{eq:kernel}
\end{equation}
where $i$ is the inclination angle of the symmetry axis, defined as
the angle between the symmetry axis and the line-of-sight; see
Fig.~\ref{fig:ellipse}. The kernel $P(x,y|R,Z)$ satisfies the
normalisation condition
\begin{equation}
  \int_{-\infty}^{\infty} \!\!\!\! dx \int_{-\infty}^{\infty} \!\!\!\! dy \;
  P(x,y|R,Z)=1. 
\end{equation}
Assuming that the orientation of the symmetry axis is known and that
one has a two-dimensional map of the projection $\psi$, one can
reverse the convolution using the iterative Richardson-Lucy
deconvolution technique \citep[see][]{LU74.1,LU94.1} and solve for
$\phi$ as a function of $R$ and $Z$. Starting with an initial guess
$\phi_0$ for $\phi$ and using the Richardson-Lucy iteration scheme,
\begin{equation}
  \phi_{n+1}(R,Z) = \phi_{n}(R,Z) \int_{-\infty}^\infty \!\!\!\! dx \int_{-\infty}^{\infty} \!\!\!\! dy \;
                    \frac{\psi(x,y)}{\psi_n(x,y)} \; P(x,y|R,Z),
\label{eq:iteration_int_xy}
\end{equation}
one can obtain approximations of $\phi$ with increasing quality. Here,
$\psi_n$ is the projection along the line-of-sight of the
approximation $\phi_n$. If we plug Eq.~(\ref{eq:kernel}) into
Eq.~(\ref{eq:iteration_int_xy}), perform the $y$ integration, and use
the new coordinate $\alpha$ defined by $x=R\cos\alpha$, we obtain
\citep[see also][]{BI90.1,RE00.1}
\begin{equation}
  \frac{\phi_{n+1}(R,Z)}{\phi_{n}(R,Z)}  = \! \int_{0}^{2\pi} \! \frac{d\alpha}{2\pi} \; 
  \frac{\psi \; (R \cos \alpha, Z \sin i + R \sin \alpha \cos i)}{\psi_n\;(R \cos \alpha, Z \sin i + R \sin \alpha \cos i)},
\label{eq:iteration_int_ellipse}
\end{equation}
where the integration follows the ellipse shown in
Fig.~\ref{fig:ellipse}.

For numerical reconstructions of axisymmetric three-dimensional
distributions, we replace the integral in
Eq.~(\ref{eq:iteration_int_ellipse}) by a sum over points, which are
distributed along the ellipse and equally spaced in $\alpha$. To
evaluate the sum, we need to find the ratio $\psi/\psi_n$ at these
points. We use two grids for the iteration, one in $x,y$-space for
$\psi$ and $\psi_n$, and one in $R,Z$-space for $\phi_n$. First, we
project $\phi_n$ along the line-of-sight on the grid in $x,y$-space to
find $\psi_n$. We do not use the kernel function for that, but perform
a direct summation using a discretised version of the first equality
in Eq.~(\ref{eq:projection}). The projection integral is approximated
by a sum over $N_z$ equally spaced points that cover a section of
length $L_z$ of the line-of-sight. This section is centred on the
$z$-coordinate of the halo. Then, $\psi_n$ is obtained by
\begin{equation}
  \psi_{n}(x_j,y_k)=\frac{L_z}{N_z} \!
  \sum_{l=1}^{N_z} \phi_n(R(x_j,y_k,z_l),Z(x_j,y_k,z_l)), 
\label{eq:projection_sum}
\end{equation}
where $x_j$ and $y_k$ are the $x$ and $y$ coordinates of the grid point $(j,k)$. The $z_l$ are the $z$ coordinates of the $N_z$ points used for the projection. The function
$\phi_n(R(x_j,y_k,z_l),Z(x_j,y_k,z_l))$ is
approximated by bilinear interpolation from the values of $\phi_n$ at
neighbouring grid points in $R,Z$-space. Since we know $\psi$ and have
calculated the projection $\psi_n$ of $\phi_n$, we can find the ratio
$\psi/\psi_n$ at points on the ellipse by bilinear interpolation from
neighbouring points of the $x,y$-space grid. This allows us to
approximate the integral in Eq.~(\ref{eq:iteration_int_ellipse}) by a
sum over $N_\alpha$ points,
\begin{equation}
  \frac{\phi_{n+1}(R,Z)}{\phi_{n}(R,Z)} = \frac{1}{N_\alpha} \! \sum_{m=1}^{N_\alpha}  
  \frac{\psi \; (R \cos \alpha_m, Z \sin i + R \sin \alpha_m \cos i)}{\psi_n\;(R \cos \alpha_m, Z \sin i + R \sin \alpha_m \cos i)},
\label{eq:iteration_sum_ellipse}
\end{equation}
and find $\phi_{n+1}$ which completes the iteration step. Here
$\alpha_m= 2 \pi m / N_\alpha$.

\subsection{Boundary effects, artifacts and regularisation}
\label{sec:regularisation}

There is, however, a problem. Assume that $L_z$ corresponds to the
height of the box shown in Fig.~\ref{fig:ellipse}, and that the area
covered by our map of $\psi$ corresponds to its top surface. To
calculate $\psi_{n+1}$ there, we have to know $\phi_{n+1}$ everywhere
in the box. But for finding $\phi_{n+1}$ close to the corners of the
box, we have to evaluate Eq.~(\ref{eq:iteration_sum_ellipse}) along
ellipses that do not fit into the top surface of the box. This means
that some of the points we have to sum over lie outside our map of
$\psi$ and $\psi_n$. As suggested by \cite{RE00.1}, we replace
$\psi/\psi_n$ for those points by its value at the closest point at
the perimeter of the map. This leads to some artifacts in the
reconstruction of $\phi$ for large $R$ and $Z$, but yields very good
results in the central region, which we are most interested in.

To start the iteration, we have to choose a guess or prior
$\phi_0$. We adopt the simplest choice of a flat or constant prior. We
set its value so as to reproduce the average value $\langle \psi
\rangle$ of the map $\psi$, namely $\phi_0 = \langle \psi \rangle /
L_z$.

\begin{figure}[ht]
\scalebox{0.45}
{
  \begin{picture}(0,0)
    \includegraphics{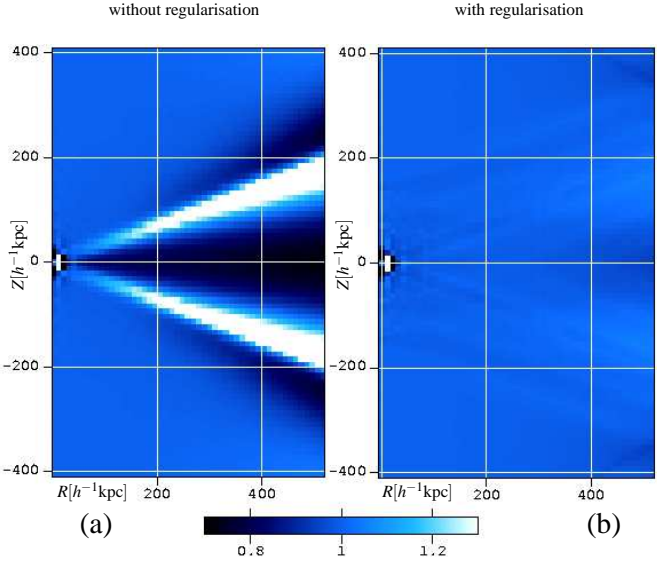}
  \end{picture}
  \setlength{\unitlength}{1mm}
  \begin{picture}(200,170)(0,0)
    \put(53,162){\makebox(0,0){\strut{}\Large without regularisation}}
    \put(26,20){\makebox(0,0){\strut{}\Large $R [h^{-1} \textrm{kpc}]$}}
    \put(4,90){\rotatebox{90}{\makebox(0,0){\strut{}\Large $Z [h^{-1} \textrm{kpc}]$}}}
    \put(27,10){\makebox(0,0){\strut{}\Huge (a)}}
    \put(152,162){\makebox(0,0){\strut{}\Large with regularisation}}
    \put(122,20){\makebox(0,0){\strut{}\Large $R [h^{-1} \textrm{kpc}]$}}
    \put(100,90){\rotatebox{90}{\makebox(0,0){\strut{}\Large $Z [h^{-1} \textrm{kpc}]$}}}
    \put(177,10){\makebox(0,0){\strut{}\Huge (b)}}
  \end{picture}
}
\caption{Ratio of the reconstructed to the original X-ray emissivity
  after $n=30$ iteration steps, $\phi_n/\phi$. The halo centre is at
  the centre left of each plot. Panel (a) shows the ratio obtained
  without regularisation. One can clearly see the spike-shaped artifacts
  of the reconstruction. Panel (b) shows the ratio for the
  reconstruction including the regularisation. It is close to unity
  everywhere except very close to the centre, where the algorithm
  converges slowly and grid resolution becomes a problem.}
\label{fig:regulate}
\end{figure}

The algorithm described above can be used to reconstruct axisymmetric
three-dimensional distributions from two-dimensional maps of its
projection along the line-of-sight. However, it runs into problems for
strongly peaked distributions such as the X-ray emissivity of a galaxy
cluster. In order to illustrate that, we reconstructed the X-ray
emissivity from an X-ray flux map, which we obtained by projecting the
emissivity of an analytically modeled, axisymmetric cluster halo. The
halo model is discussed in Sect.~\ref{sec:analytic_halos}. For the
projection, we chose an inclination angle of $i=70^\circ$. We
performed the reconstruction with a rather high number of $n=30$
iterations. In the left panel of Fig.~\ref{fig:regulate}, we show the
ratio between the reconstructed and the original X-ray emissivity. One
can clearly see spike-shaped artifacts of the reconstruction. The angle
between these spikes and the symmetry axis is equal to the inclination
$i$. This means that the ellipses corresponding to $R$ and $Z$ values
of points in the spikes pass directly through the halo centre in the
map of $\psi$.

Richardson-Lucy deconvolution reproduces large structures quickly,
while it converges slowly to small scale structures such as the peak
at the halo centre \citep[see][]{LU74.1,LU94.1}. This means that, when
starting with a flat prior, $\psi/\psi_n$ can be quite large close to
the centre even after several iterations. Thus, when we evaluate
(\ref{eq:iteration_sum_ellipse}) for points further out whose ellipses
pass through the halo centre, we find ratios of
$\phi_{n+1}(R,Z)/\phi_{n}(R,Z)$ which are too high, and the spike-shaped
artifacts form. They appear already after the first few iterations and
are very stable. In the left panel of Fig.~\ref{fig:regulate}, we show
them after 30 iterations, and it would take several hundred more
iterations until they slowly disappear.

To prevent the formation of such artifacts, we use a regularisation
scheme. First, we calculate an average $\langle \psi/\psi_n \rangle$
for the points used in the sum in
Eq.~(\ref{eq:iteration_sum_ellipse}), which is defined by
\begin{align}
  \langle \psi/\psi_n \rangle (R,Z) & \equiv \nonumber \\
  \equiv \frac{1}{N_\alpha} \! \sum_{m=1}^{N_\alpha} &
  \textrm{min}\,\Big(\frac{\psi\,(R \cos \alpha_m,\, Z \sin i + R \sin \alpha_m \cos i)}
  {\psi_n(R \cos \alpha_m,\, Z \sin i + R \sin \alpha_m \cos i)},10\,\Big).
  \label{eq:reg_it}
\end{align}
Then we set
\begin{align}
  c_n(R,Z) \equiv \frac{1}{N_\alpha} \! \sum_{m=1}^{N_\alpha} \textrm{min}\,\Big(& 
  \frac{\psi \, (R \cos \alpha_m,\, Z \sin i + R \sin \alpha_m \cos i)}{\psi_n(R \cos \alpha_m,\, Z \sin i + R \sin \alpha_m \cos i)}, \nonumber \\
  & \quad 1.2 \, \langle \psi/\psi_n \rangle (R,Z)\,\Big),
  \label{eq:reg_it2}   
\end{align}
and use 
\begin{equation}
  \frac{\phi_{n+1}(R,Z)}{\phi_{n}(R,Z)} = \textrm{max}\,(\,c_n(R,Z),\,0.25\;),
  \label{eq:reg_it3}  
\end{equation}
to calculate $\phi_{n+1}(R,Z)$. This regularisation of the iteration
scheme supresses the formation of the spike-shaped artifacts. It limits
the impact of sharp peaks in $\psi$ on points that are far away from
the corresponding peaks in $\phi$ by using an upper limit for
$\psi/\psi_n$ in Eqs. (\ref{eq:reg_it}) and (\ref{eq:reg_it2}). The effectiveness
of the regularisation scheme is not very sensitive to the exact
numerical values of the upper limits, which we chose by trial and
error, as long as they suppress sharp peaks and are not too
restrictive to allow convergence in a reasonable number of
iterations. The lower limit for the correction factor in
Eq.~(\ref{eq:reg_it3}) is introduced just to make sure that
$\phi_{n+1}$ does not change its sign or become very small in the
first few iteration steps, which would potentially cause problems
later.

We repeated the reconstruction of the X-ray emissivity using this
regularisation. In the right panel of Fig.~\ref{fig:regulate}, we show
the ratio of the reconstructed to the original emissivity after 30
iterations. The spikes that are present in the left panel have almost
disappeared. The ratio is close to unity everywhere in the region
shown, except very near the halo centre where grid resolution and the
slow convergence to small-scale structures becomes a problem. Apart
from that, the deprojection works very well. The errors are usually
smaller than $1\%$.

\subsection{Inclination angle}
\label{sec:inc_simple}

So far, we have assumed that the orientation of the symmetry axis is
known. In reality, when applying this algorithm to observations, this
will not be the case. However, the orientation of the symmetry axis in
the plane of the sky can directly be obtained from the map $\psi$. Its
inclination $i$ on the other hand can be found in the following
way. First, we repeat the reconstruction as described above, assuming
several different plausible values of $i$ and using a fixed
number $N_{\textrm{it}}$ of iterations. Then, we compare the maps
$\psi_{N_{\textrm{it}}}$, corresponding to the reconstructed
distributions, to the original (or observed) map $\psi$ and find the
value of $i$ for which it fits best. This can for example be done by
minimising the penalty function
\begin{equation}
  \sum_{j,k}\frac{(\psi(x_j,y_k)-\psi_{N_{\textrm{it}}}(x_j,y_k))^2}{\psi^2(x_j,y_k)},
\label{eq:penalty_function}     
\end{equation}
where, depending on the shape of the distribution one wants to
reconstruct, it may be favourable to sum only over points in the
central region of the map.

\subsection{Reconstruction of density and temperature of the ICM from combined X-ray and
  thermal Sunyaev-Zel'dovich effect observations}
\label{sec:depro_rho_temp}

So far, we have discussed how to reconstruct a single
three-dimensional distribution of a physical quantity from a single
two-dimensional map of its projection along the
line-of-sight. However, one can obtain additional information by
combining maps from different observations \citep[see
also][]{RE01.1,RE00.1}. Here, we reconstruct several physical
quantities at the same time by combining different observations that
depend on these quantities. Specifically, we will show how to obtain
three-dimensional distributions of the density and temperature of the
intra-cluster medium (ICM) in axisymmetric cluster halos by combining
observations of the (thermal) X-ray emission and the (thermal)
Sunyaev-Zel'dovich effect. The X-ray surface brightness is taken to be
proportional to
\begin{equation}
  \psi_{\textrm{X-ray}} \equiv \int dz \, \rho^2 \, \lambda(T,\cal{Z}),
\label{eq:xray_proj}
\end{equation}
where $\rho$ and $T$ are the gas density and temperature,
respectively. The integral extends along the line-of-sight. The
cooling function $\lambda(T,\cal{Z})$ depends on the temperature and
the metallicity $\cal{Z}$, and $\cal{Z}$ is assumed to be constant
here. The (thermal) Sunyaev-Zel'dovich decrement or increment is
proportional to the Compton $y$-parameter 
\begin{equation}
  y=\int \frac{k_B T}{m_e c^2} n_e \sigma_T c d t,
\end{equation}
where the integration also follows the line-of-sight,
$k_B$ is Boltzmann's constant, $m_e$ the electron mass, $c$ the speed of
light, $\sigma_T$ the Thomson cross section, and $n_e$ the electron
number density at the position passed by the light ray at time
$t$. For fully ionised gas with constant metallicity, this is also
proportional to $\psi_{\textrm{SZ}}$ defined by
\begin{equation}
  \psi_{\textrm{SZ}} \equiv \int dz \, \rho \, T.
\label{eq:sz_proj}
\end{equation}
$\psi_{\textrm{X-ray}}$ and $\psi_{\textrm{SZ}}$ can be obtained from
observations. For reconstructing the ICM temperature and density, we
start from some initial guess $\rho_0(R,Z)$ and $T_0(R,Z)$. In analogy
to Eq.~(\ref{eq:projection_sum}), we use discrete approximations of Eqs.
(\ref{eq:xray_proj}) and (\ref{eq:sz_proj}),
\begin{align}
  \psi_{\textrm{X-ray},\,n}(x_j,&y_k)=\frac{L_z}{N_z} \!
  \sum_{l=1}^{N_z} \rho_n^2(R(x_j,y_k,z_l),Z(x_j,y_k,z_l)) 
  \nonumber \\
  &\times \, \lambda(T(R(x_j,y_k,z_l),Z(x_j,y_k,z_l)),\cal{Z}), \\
  \psi_{\textrm{SZ},\,n}(x_j,&y_k)=\frac{L_z}{N_z} \!
  \sum_{l=1}^{N_z} \rho_n(R(x_j,y_k,z_l),Z(x_j,y_k,z_l)) 
  \nonumber \\
  &\times \, T(R(x_j,y_k,z_l),Z(x_j,y_k,z_l)),
\end{align}
to obtain $\psi_{\textrm{X-ray},\,0}$ and $\psi_{\textrm{SZ},\,0}$. In
analogy to Eq.~(\ref{eq:iteration_sum_ellipse}), we define
\begin{align}
  c_{\textrm{X-ray},n}(R,Z) &= \! \frac{1}{N_\alpha} \! \sum_{m=1}^{N_\alpha}  
  \frac{\psi_{\textrm{X-ray}} (R \cos \alpha_m, Z \sin i + \! R \sin \alpha_m \cos i)}{\psi_{\textrm{X-ray},n}(R \cos \alpha_m, Z \sin i + \! R \sin \alpha_m \cos i)}, 
  \label{eq:corr_xray} \\
  c_{\textrm{SZ},\,n}(R,Z) &= \! \frac{1}{N_\alpha} \! \sum_{m=1}^{N_\alpha}  
  \frac{\psi_{\textrm{SZ}} (R \cos \alpha_m, Z \sin i + \! R \sin \alpha_m \cos i)}{\psi_{\textrm{SZ},\,n}(R \cos \alpha_m, Z \sin i + \! R \sin \alpha_m \cos i)},
  \label{eq:corr_sz}
\end{align}
and use the iteration scheme
\begin{align}
  \frac{\rho^2_{n+1}\,\lambda(T_{n+1},\cal{Z})}{\rho^2_{n}\,\lambda(T_{n},\cal{Z})} &= c_{\textrm{X-ray},\;n}, 
  \label{eq:iteration_xray} \\
  \frac{\rho_{n+1} T_{n+1}}{\rho_{n} T_{n}} &= c_{\textrm{SZ},\;n}.
  \label{eq:iteration_sz}
\end{align}
In order to find the next iterative approximation for the density and
the temperature, we have to solve equations (\ref{eq:iteration_xray})
and (\ref{eq:iteration_sz}) for $\rho_{n+1}$ and $T_{n+1}$. To include
line emission in the cooling function, one could tabulate
$\lambda(T_{n},\cal{Z})$, e.g.~using the software package XSPEC \citep[see e.g.][]{AR96.1} for a
specific emission model and metallicity, and solve the equations above
numerically. However, since the main focus of this paper is to
demonstrate the deprojection algorithm, we use the simple assumption
of continuous bremsstrahlung,
\begin{equation}
  \lambda(T,{\cal Z})=\sqrt{T}.
\end{equation}
Then, we obtain from (\ref{eq:iteration_xray}) and
(\ref{eq:iteration_sz}),
\begin{align}
  \rho_{n+1}=\frac{c_{\textrm{X-ray},\,n}^{2/3}}{c_{\textrm{SZ},\,n}^{1/3}}\rho_n,\\
  T_{n+1}=\frac{c_{\textrm{SZ},\,n}^{4/3}}{c_{\textrm{X-ray},\,n}^{2/3}}T_n.
\end{align}
Note that for evaluating Eq.~(\ref{eq:corr_xray}) and
(\ref{eq:corr_sz}), we also use the regularisation introduced in
Sect.~\ref{sec:regularisation} and expressed by
Eqs.~(\ref{eq:reg_it})-(\ref{eq:reg_it3}). Again, for points that lie
outside the map of $\psi_{\textrm{X-ray}}$ and $\psi_{\textrm{SZ}}$,
the ratios $\psi_{\textrm{X-ray}}/\psi_{\textrm{X-ray},\,n}$ and
$\psi_{\textrm{SZ}}/\psi_{\textrm{SZ},\,n}$ are approximated by their
values at the closest point at the perimeter of the map. In the next
two sections, we shall apply this deprojection algorithm to
axisymmetric analytic halos and to numerically simulated cluster
halos, and discuss its performance.

\section{Deprojection of analytic halos}
\label{sec:analytic_halos}

The analytic halo model which we use to test the deprojection
algorithm is based on the NFW density profile. However, for the
deprojection to be non-trivial, we prefer to have ellipsoidal
iso-density surfaces. We therefore assume that the density of the ICM
is a function of
\begin{equation}
  r \equiv \sqrt{\frac{R^2}{R_s^2}+\frac{Z^2}{Z_s^2}},
\end{equation}
where $R_s$ is a scaling radius perpendicular to the symmetry axis,
and $Z_s$ is a scaling distance along the axis. The density of the hot
cluster gas is then taken to be
\begin{equation}
  \rho = \frac{\rho_s}{(\epsilon_r+r)(1+r)^2}, 
\end{equation}
which differs from an NFW halo not only by its ellipsoidal shape but
also by the small constant $\epsilon_r = 0.001$ introduced to ensure
that the density does not diverge for $r\rightarrow0$. This divergence
would cause problems in numerical calculations. For the gas
temperature we use a phenomenological description that roughly
corresponds to the temperature profiles found in the simulated cluster sample that we will use in Sect. \ref{sec:num_halos}. Namely, we set
\begin{equation}
  T = T_{\textrm{max}}r^{-0.2\gamma(r)},
\end{equation}
where $\gamma(r)=\tanh(3(r-1))$ is $-1$ for $r \ll 1$ and $+1$ for $r
\gg 1$. The values of the parameters $\rho_s = 3\times 10^5 h^{-1}
M_\odot / (h^{-1} \textrm{kpc})^3$, $Z_s = 500 h^{-1} \textrm{kpc}$,
$R_s = 300 h^{-1} \textrm{kpc}$, and $T_\textrm{max} = 12
\textrm{keV}$, which we use correspond to the gas component of a
massive galaxy cluster. Here, $h$ is the reduced Hubble constant,
which we set to $0.7$.

\subsection{Deprojection of analytic halos without observational
  noise}

\begin{figure}[ht]
\scalebox{0.45}
{
  \begin{picture}(0,0)
    \includegraphics{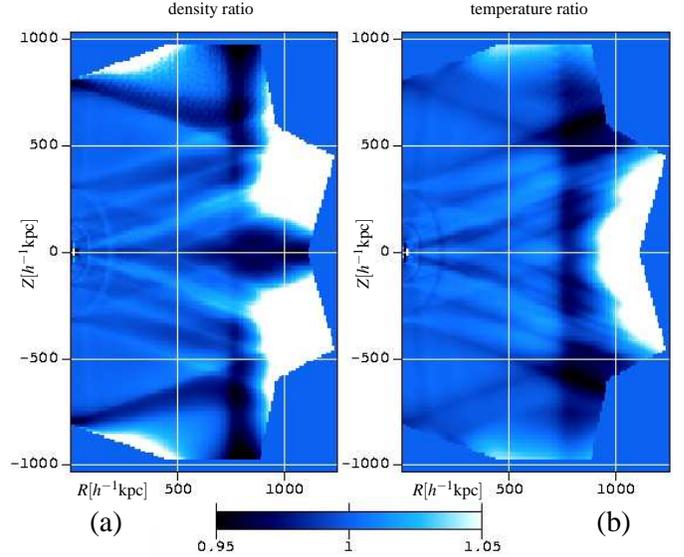}
  \end{picture}
  \setlength{\unitlength}{1mm}
  \begin{picture}(200,170)(0,0)
    \put(58,162){\makebox(0,0){\strut{}\Large density ratio}}
    \put(29,20){\makebox(0,0){\strut{}\Large $R [h^{-1} \textrm{kpc}]$}}
    \put(4,90){\rotatebox{90}{\makebox(0,0){\strut{}\Large $Z [h^{-1} \textrm{kpc}]$}}}
    \put(27,10){\makebox(0,0){\strut{}\Huge (a)}}
    \put(152,162){\makebox(0,0){\strut{}\Large temperature ratio}}
    \put(128,20){\makebox(0,0){\strut{}\Large $R [h^{-1} \textrm{kpc}]$}}
    \put(103,90){\rotatebox{90}{\makebox(0,0){\strut{}\Large $Z [h^{-1} \textrm{kpc}]$}}}
    \put(177,10){\makebox(0,0){\strut{}\Huge (b)}}
\end{picture}
}
\caption{Ratio of the reconstructed to the original density and
  temperature for the analytic halo after $n=20$ iterations. An
  inclination angle of $i=70^\circ$ was chosen and assumed to be known
  in performing the reconstruction. Panel (a) shows the density ratio
  $\rho_n/\rho$ and panel (b) the temperature ratio $T_n/T$. In the
  central region, the errors are of the order of $1\%$. We plot the
  ratios for all $R$ and $Z$ values possible within the box used for
  the reconstruction (see Fig.~\ref{fig:ellipse}). This causes the
  star-like shape of the perimeter of the plot in the $R$, $Z$
  plane. Close to that boundary, at large $R$ or $Z$ values, the
  ratios can significantly differ from unity. This is, however,
  expected because the ellipses used in the reconstruction
  of $\rho$ and $T$ at those points lie mostly outside the maps of
  $\psi_{\textrm{X-ray}}$ and $\psi_{\textrm{SZ}}$.}
\label{fig:RZratio_NFW_70}
\end{figure}

Having chosen an inclination angle $i$, we project the analytic halo
described above on a $128\times128$ grid with a sidelength of $1.5
h^{-1} \textrm{Mpc}$ and obtain the X-ray and Sunyaev-Zel'dovich
effect maps, $\psi_{\textrm{X-ray}}$ and $\psi_{\textrm{SZ}}$. We use
the algorithm discussed in Sect.~\ref{sec:depro_rho_temp} with these
maps to reconstruct the gas density and temperature. In
Fig.~\ref{fig:RZratio_NFW_70}, we compare the results of the
deprojection with the original density and temperature of the analytic
halo for an inclination angle $i=70^\circ$ and after $n=20$
iterations. The inclination was assumed to be known in performing the
deprojection. The left and right panels show the density and
temperature ratios, $\rho_n/\rho$ and $T_n/T$, repsectively. The
star-like pattern of the plots maps the ranges of the $R$ and $Z$
coordinates occurring in the simulation box used for the
reconstruction (see Fig.~\ref{fig:ellipse}). In the central region of
the cluster, the reconstruction works very well. Errors are of the
order of $1\%$. Despite the regularisation, one can still see some
remains of the spike-shaped artifacts discussed in
Sect.~\ref{sec:regularisation}. For large $R$ or $Z$ values, close to
the star-shaped boundary of the plots, the quality of the
reconstruction decreases. This is not at all surprising because the
ellipses along which Eqs.~(\ref{eq:corr_xray}) and (\ref{eq:corr_sz})
must be evaluated to reconstruct density and temperature at those
points lie mostly outside of the maps of $\psi_{\textrm{X-ray}}$ and
$\psi_{\textrm{SZ}}$.

\begin{figure}[ht]
\scalebox{0.7}{
\begin{picture}(0,0)%
\includegraphics{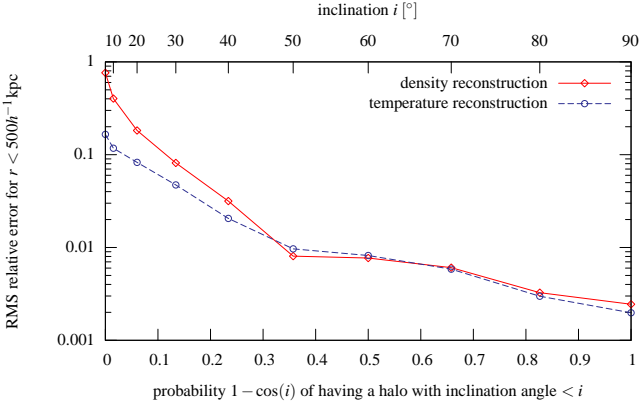}%
\end{picture}%
\begingroup
\setlength{\unitlength}{0.0200bp}%
\begin{picture}(18000,10800)(0,0)%
\put(2750,1650){\makebox(0,0)[r]{\strut{} 0.001}}%
\put(2750,4150){\makebox(0,0)[r]{\strut{} 0.01}}%
\put(2750,6650){\makebox(0,0)[r]{\strut{} 0.1}}%
\put(2750,9150){\makebox(0,0)[r]{\strut{} 1}}%
\put(3025,1100){\makebox(0,0){\strut{} 0}}%
\put(4440,1100){\makebox(0,0){\strut{} 0.1}}%
\put(5855,1100){\makebox(0,0){\strut{} 0.2}}%
\put(7270,1100){\makebox(0,0){\strut{} 0.3}}%
\put(8685,1100){\makebox(0,0){\strut{} 0.4}}%
\put(10100,1100){\makebox(0,0){\strut{} 0.5}}%
\put(11515,1100){\makebox(0,0){\strut{} 0.6}}%
\put(12930,1100){\makebox(0,0){\strut{} 0.7}}%
\put(14345,1100){\makebox(0,0){\strut{} 0.8}}%
\put(15760,1100){\makebox(0,0){\strut{} 0.9}}%
\put(17175,1100){\makebox(0,0){\strut{} 1}}%
\put(17175,9700){\makebox(0,0){\strut{}90}}%
\put(14718,9700){\makebox(0,0){\strut{}80}}%
\put(12335,9700){\makebox(0,0){\strut{}70}}%
\put(10100,9700){\makebox(0,0){\strut{}60}}%
\put(8080,9700){\makebox(0,0){\strut{}50}}%
\put(6335,9700){\makebox(0,0){\strut{}40}}%
\put(4921,9700){\makebox(0,0){\strut{}30}}%
\put(3878,9700){\makebox(0,0){\strut{}20}}%
\put(3240,9700){\makebox(0,0){\strut{}10}}%
\put(550,5400){\rotatebox{90}{\makebox(0,0){\strut{}RMS relative error for $r < 500 h^{-1} \textrm{kpc}$}}}%
\put(10100,275){\makebox(0,0){\strut{}probability $1-\cos(i)$ of having a halo with inclination angle $<i$}}%
\put(10100,10524){\makebox(0,0){\strut{}inclination $i \; [^\circ]$}}%
\put(14950,8575){\makebox(0,0)[r]{\strut{}density reconstruction}}%
\put(14950,8025){\makebox(0,0)[r]{\strut{}temperature reconstruction}}%
\end{picture}%
\endgroup
}
\caption{Dependence of the quality of the deprojection of analytic
  halos on the inclination of the symmetry axis. We show the
  volume-weighted RMS relative errors
  $(\rho_n-\rho)/\rho$ and $(T_n-T)/T$ as functions of the inclination
  angle $i$ after $n=20$ iterations. The averages were computed within
  a sphere of radius $500 h^{-1} \textrm{kpc}$ around the halo
  centre. The quantity $1-\cos(i)$ shown on the abscissae is chosen
  such as to have a flat number-density distribution for randomly
  oriented halos. The inclination angle $i$ was assumed to be known in
  performing the deprojection.}
\label{fig:haloinc_quality}
\end{figure}

Note that the quality of the reconstruction also depends on the
inclination of the halo's symmetry axis. Of course, best results are
achieved when the symmetry axis is perpendicular to the
line-of-sight. Then the assumption of axial symmetry contains the most
information. If, on the other hand, the symmetry axis is parallel to
the line-of-sight, the axial symmetry just corresponds to the circular
symmetry of the maps $\psi_{\textrm{X-ray}}$ and $\psi_{\textrm{SZ}}$
and does not yield any useful additional
information. Figure~\ref{fig:haloinc_quality} illustrates this
inclination dependence. It shows the volume-weighted root mean square
(RMS) relative errors of the reconstructed gas density and
temperature, computed within a sphere of radius $500 h^{-1}
\textrm{kpc}$ around the halo centre. Again, the knowledge of the
inclination angle $i$ was used in the deprojection. An accuracy of
$1\%$ or better is achieved for about two thirds of the analytic halos
in a randomly oriented sample. However, halos that happen to have a
very small inclination angle $i$ are necessarily poorly reconstructed.

\subsection{Deprojection of analytic halos including observational
  noise}
\label{sec:analytic_halos_noise}

So far we have not considered noise that will be present in any real
X-ray or Sunyaev-Zel'dovich effect observation. We will now discuss
the impact it has on the reconstruction of ICM densities and
temperatures.

We model the noise in X-ray observations as follows. First, we
calculate for each pixel $(j,k)$ of the halo's X-ray map
$\psi_{\textrm{X-ray}}$ the number of photons $\langle
N_{\gamma\,j,k}\rangle$ expected from \emph{bremsstrahlung}, which is
proportional to $\langle
N_{\gamma\,j,k}\rangle\sim\sum_{l=1}^{N_Z}\textrm{E}_1(\frac{E_{\textrm{min}}}{k_B T})\frac{\rho^2}{\sqrt{T}}$,
where the sum extends along the line-of-sight represented by the pixel $(j,k)$, and $\textrm{E}_1$ is the exponential
integral function. $E_{\textrm{min}}$ is a lower energy cutoff which
is necessary because the number of photons emitted is infrared
divergent. We choose $E_{\textrm{min}}=0.23\,\textrm{keV}$, which is a
reasonable lower limit for the photons from galaxy clusters observed
in current X-ray experiments. Next, we normalise the numbers of
expected photons such that they sum up to $\sum_{j,k}\langle
N_{\gamma\,j,k}\rangle=10^4$ on the entire map. For each pixel $(j,k)$,
we then set the actual number of photon counts $N_{\gamma\,j,k}$ to a
value drawn from a Poisson distribution with expectation value
$\langle N_{\gamma\,j,k}\rangle$. We then add the noise to the map
$\psi_{\textrm{X-ray}}$ by multiplying $\psi_{\textrm{X-ray}\,j,k}$
with $N_{\gamma\,j,k}/\langle N_{\gamma\,j,k}\rangle$ for all pixels.

For the Sunyaev-Zel'dovich effect, we add noise corresponding to
future ALMA Band 3 observations \citep[see][]{BU99.1}. In Band 3 (84
to 116 GHz) and in its compact configuration, ALMA will be able to
achieve a temperature sensitivity of $50 \mu\textrm{K}$ at a spatial
resolution of $\sim 3$ arcsec in about four hours of observation. At
an assumed halo redshift of $0.3$, this resolution corresponds to the
angular size chosen for the pixels of the map $\psi_{\textrm{SZ}}$. We
convert the temperature sensitivity cited above to an error
$\sigma_{\psi_{\textrm{SZ}}}$ of $\psi_{\textrm{SZ}}$. Then, for each
pixel, we add noise obtained from a normal distribution with standard
deviation $\sigma_{\psi_{\textrm{SZ}}}$ to $\psi_{\textrm{SZ}}$.

Richardson-Lucy deconvolution has the nice property of approximating
large-scale features quickly and small-scale noise slowly. Yet, it
turns out that smoothing the noisy maps $\psi_{\textrm{X-ray}}$ and
$\psi_{\textrm{SZ}}$ before using them in the deprojection improves
the results considerably. We use the following smoothing scheme. For
the X-ray observations, we assume that in addition to the map
$\psi_{\textrm{X-ray}}$ we also know the photon counts
$N_{\gamma\,j,k}$ for all pixels. We then calculate for each pixel
$(j,k)$ a radius $h_{j,k}$ so that we have a fixed number of 100 photons
inside a circle with radius $h_{j,k}$ around that pixel. After that we
redistribute the value $\psi_{\textrm{X-ray}\,j,k}$ of each pixel on
the grid with a smoothing kernel of width $h_{j,k}$ centred on that
pixel. This greatly reduces the fluctuations in the map
$\psi_{\textrm{X-ray}}$ caused by photon noise. In the remainder of the paper we will call this first step of the smoothing scheme ``photon-noise smoothing''.  

For the smoothing kernel, we take the line-of-sight projection of the
cubic spline SPH kernel $W(r,h)$ defined in Appendix A of
\cite{SP01.1}. It is well suited for this purpose and allows us to use
the same routine for smoothing here and for projecting the numerical
SPH halos we use in Sect.~\ref{sec:num_halos}. For axisymmetric halos,
the projection should be symmetric about the projected axis. However, the symmetry is broken here by noise. We restore it before performing the deprojection. Since we have oriented the grid
such that it is parallel to and centred on the projected symmetry axis, we can do that by replacing
$\psi_{\textrm{X-ray}\,j,k}$ and
$\psi_{\textrm{X-ray}\,N_\textrm{grid}-j,k}$ by their arithmetic
mean. Here, $N_\textrm{grid}=128$ is the dimension of the grid. We
symmetrise $\psi_{\textrm{SZ}}$ in the same way. After that we perform
one more smoothing operation on $\psi_{\textrm{X-ray}}$ and
$\psi_{\textrm{SZ}}$ to further reduce fluctuations caused by
noise.

In numerically simulated halos, which we will discuss later, this will
also suppress the effect of subclumps. Since we do not want to smooth
out the peaks in the halo core, we choose a smoothing length $h$ that
depends on the distance $r$ from the halo (or map) centre, namely $h =
h_\textrm{max} ( 1 - W(r,r_\textrm{max})/W(0,r_\textrm{max}) )$. It is
zero in the centre of the map and continually increases to $h = 375
h^{-1}\textrm{kpc}$ at a radius equal to $r_\textrm{max} = 750
h^{-1}\textrm{kpc}$ or larger. This yields smallest RMS errors. Once we have
calculated $h$ for each pixel, we smooth the maps of
$\psi_{\textrm{X-ray}}$ and $\psi_{\textrm{SZ}}$ with the projection
of the SPH smoothing kernel mentioned above and with the
position-dependent smoothing length $h$. Note that roughly 80\% of a pixel's value is redistributed within a circle of radius $h/2$. We refer to this second step of the smoothing scheme as ``radius-dependent smoothing''.

\begin{figure}[ht]
\scalebox{0.45}
{
  \begin{picture}(0,0)
    \includegraphics{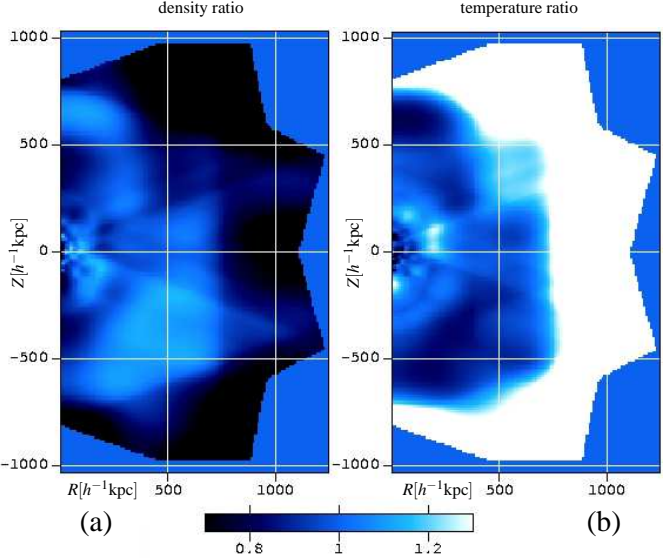}
  \end{picture}
  \setlength{\unitlength}{1mm}
  \begin{picture}(200,170)(0,0)
    \put(58,162){\makebox(0,0){\strut{}\Large density ratio}}
    \put(29,20){\makebox(0,0){\strut{}\Large $R [h^{-1} \textrm{kpc}]$}}
    \put(4,90){\rotatebox{90}{\makebox(0,0){\strut{}\Large $Z [h^{-1} \textrm{kpc}]$}}}
    \put(27,10){\makebox(0,0){\strut{}\Huge (a)}}
    \put(152,162){\makebox(0,0){\strut{}\Large temperature ratio}}
    \put(128,20){\makebox(0,0){\strut{}\Large $R [h^{-1} \textrm{kpc}]$}}
    \put(103,90){\rotatebox{90}{\makebox(0,0){\strut{}\Large $Z [h^{-1} \textrm{kpc}]$}}}
    \put(177,10){\makebox(0,0){\strut{}\Huge (b)}}
  \end{picture}
}
\caption{Ratio of the reconstructed to the original density and
  temperature for the analytic halo with observational noise after
  $n=5$ iterations. An inclination of $i=70^\circ$ was chosen and
  assumed to be known in performing the reconstruction. ``Photon-noise'' and ``radius-dependent'' smoothing were applied. Panel (a) shows the density ratio $\rho_n/\rho$, and panel (b) the temperature
  ratio $T_n/T$. In the central region, the errors are of the order of
  $5\%$ to $10\%$.}
\label{fig:RZratio_noise_NFW_70}
\end{figure}

After degrading the analytic halo with noise and applying the
smoothing scheme described above, we perform the iterative
deprojection. The results after $n=5$ iterations are shown in
Fig.~\ref{fig:RZratio_noise_NFW_70}. An inclination of $i=70^\circ$
was chosen and assumed to be known in the deprojection. Again, the
left panel shows the ratio of the reconstructed to the original
density, and the right panel the corresponding temperature
ratio. Average errors in the central region are of the order of $5\%$
to $10\%$. As expected, further outside, where the signal-to-noise
ratio becomes small and the ellipses used for the reconstruction lie
mostly outside the maps of $\psi_{\textrm{X-ray}}$ and
$\psi_{\textrm{SZ}}$, the errors are substantially larger. Note that
at locations where we obtain a too low density, we usually find a too
high temperature and vice versa. This happens because the algorithm
minimises the deviations of the reconstructed from the original X-ray
and SZ-effect maps.

\begin{figure}[ht]
\scalebox{0.7}
{
  \begin{picture}(0,0)%
    \includegraphics{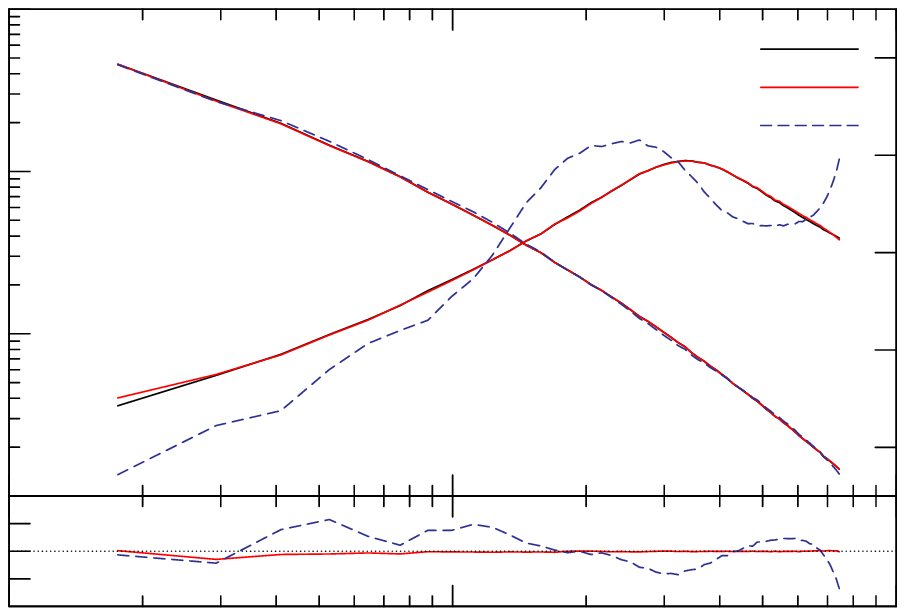}%
  \end{picture}%
    \begingroup
    \setlength{\unitlength}{0.0200bp}%
  \begin{picture}(18000,10800)(0,0)%
    \put(2750,5577){\makebox(0,0)[r]{\strut{} 1e-05}}%
    \put(2750,7913){\makebox(0,0)[r]{\strut{} 1e-04}}%
    \put(2750,10250){\makebox(0,0)[r]{\strut{} 0.001}}%
    \put(3025,2690){\makebox(0,0){\strut{}}}%
    \put(9413,2690){\makebox(0,0){\strut{}}}%
    \put(15800,2690){\makebox(0,0){\strut{}}}%
    \put(16075,3941){\makebox(0,0)[l]{\strut{} 6}}%
    \put(16075,5343){\makebox(0,0)[l]{\strut{} 8}}%
    \put(16075,6745){\makebox(0,0)[l]{\strut{} 10}}%
    \put(16075,8147){\makebox(0,0)[l]{\strut{} 12}}%
    \put(16075,9549){\makebox(0,0)[l]{\strut{} 14}}%
    \put(550,6745){\rotatebox{90}{\makebox(0,0){\strut{} gas density $[10^{10}h^2 M_\odot\textrm{kpc}^{-3}]$}}}%
    \put(17449,6745){\rotatebox{90}{\makebox(0,0){\strut{} gas temperature [keV]}}}%
    \put(13575,9675){\makebox(0,0)[r]{\strut{}original halo}}%
    \put(13575,9125){\makebox(0,0)[r]{\strut{}halo reconstructed without noise}}%
    \put(13575,8575){\makebox(0,0)[r]{\strut{}halo reconstructed with noise}}%
    \put(2750,2047){\makebox(0,0)[r]{\strut{} 0.95}}%
    \put(2750,2445){\makebox(0,0)[r]{\strut{} 1}}%
    \put(2750,2842){\makebox(0,0)[r]{\strut{} 1.05}}%
    \put(3025,1100){\makebox(0,0){\strut{} 10}}%
    \put(9413,1100){\makebox(0,0){\strut{} 100}}%
    \put(15800,1100){\makebox(0,0){\strut{} 1000}}%
    \put(600,2445){\rotatebox{90}{\makebox(0,0){\strut{} density}}}%
    \put(1000,2445){\rotatebox{90}{\makebox(0,0){\strut{} ratio }}}
    \put(1400,2445){\rotatebox{90}{\makebox(0,0){\strut{} $\rho_{n}/\rho$}}}
    \put(9412,275){\makebox(0,0){\strut{} radius $[h^{-1} \textrm{kpc}]$}}%
\end{picture}%
\endgroup
}
\caption{Density and temperature profiles of the original and the
  reconstructed analytic halos. The upper panel shows the density
  (falling curves, left axis) and the temperature profiles (rising
  curves, right axis) of the original analytic halo, the halo
  reconstructed without observational noise (and without any
  smoothing), and the halo reconstructed from maps with
  observational noise to which the complete smoothing scheme was applied.
  The lower panel shows the profile of the ratio of the reconstructed density $\rho_n$ to the original density
  $\rho$. The number of iterations used was $n=20$ in the case without
  noise and $n=5$ in the case with noise. An inclination of
  $i=70^\circ$ was chosen and assumed to be known in the
  reconstruction.}
\label{fig:analytic_profiles}
\end{figure}

In Fig.~\ref{fig:analytic_profiles}, we show density and temperature
profiles of the original analytic halo, of the halo reconstructed from
maps without observational noise, and of the halo reconstructed from
smoothed maps which contain observational noise. The reconstructed
halos are the same as shown in Figs.~\ref{fig:RZratio_NFW_70} and
\ref{fig:RZratio_noise_NFW_70}. Without noise, both the temperature
and the density profiles are reproduced very well. With noise, we can
still reproduce density profiles with an accuracy of a few
percent. The errors in the temperature profile are somewhat
larger. Deviations are mainly caused by the noise, but some are also
artifacts of the smoothing scheme we apply. Especially the too high
temperature near $\sim 200 \, h^{-1}$ kpc and the too low temperature
near $\sim 400 \, h^{-1}$ kpc are a consequence of ``radius-dependent smoothing''. On the other hand, without such
smoothing we would approximately double the errors in the density and
temperature reconstructions.

\section{Deprojection of numerically simulated cluster halos}
\label{sec:num_halos}

So far, we have demonstrated the performance of the algorithm with
axisymmetric analytic halos. We were able to reconstruct their
three-dimensional density and temperature distributions from synthetic
X-ray and (thermal) Sunyaev-Zel'dovich effect observations. Real
galaxy clusters, however, are hardly perfectly axisymmetric. We will
study in this section whether they nonetheless allow accurate density
and temperature reconstructions with the deprojection algorithm
proposed in Sect.~\ref{sec:depro_rho_temp}. We use a sample of four
numerically simulated galaxy clusters to investigate into this
question.

The simulations were carried out by Klaus Dolag with GADGET-2
\citep[]{SP05.1}, a new version of the parallel TreeSPH simulation
code GADGET \citep[]{SP01.1}. The cluster regions were extracted from
a dissipation-less (dark matter only) simulation with a box size of
$479h^{-1}$ Mpc of a flat $\Lambda$CDM model with $\Omega_m=0.3$,
$h=0.7$, $\sigma_8=0.9$ \citep[see][]{YO01.1}. They were re-simulated
with higher resolution using the ``Zoomed Initial Conditions'' (ZIC)
technique \citep[]{TO97.1}. Gas was introduced into the
high-resolution region by splitting each parent particle into a gas
and a dark matter particle, which were then displaced by half the mean
inter-particle distance, such that the centre-of-mass and the momentum
were conserved. The mass ratio of gas to dark matter particles was set
to obtain $\Omega_b=0.04$. The final mass resolution was $m_{\rm
DM}=1.13\times 10^9\:h^{-1}M_\odot$ and $m_{\rm gas}=1.7\times
10^8\:h^{-1}M_\odot$ for dark-matter and gas particles within the
high-resolution region, respectively. The simulations we use follow
the dynamics of the dark matter and the adiabatic evolution of the
cluster gas, but they ignore radiative cooling. They are described in
more detail in \cite{PU05.1} and \cite{DO05.1}.

Our deprojection algorithm requires a symmetry axis, which real
clusters do not generally have. We thus need to choose an axis around
which the numerical clusters have at least a high degree of
symmetry. We do this by calculating the inertial tensor of the cluster
gas inside a sphere of radius $500 h^{-1}$ kpc around the cluster
centre and finding its eigenvectors $\vec{v_1},\vec{v_2},\vec{v_3}$
and eigenvalues $e_1 \geq e_2 \geq e_3$. We choose the symmetry axis
through the cluster centre and parallel to the eigenvector $\vec{v_3}$
with the smallest eigenvalue if $e_1 / e_2 \leq e_2 / e_3$, or
parallel to the eigenvector $\vec{v_1}$ with the largest eigenvalue
otherwise. This means that, if two eigenvalues are very similar, we
choose the axis parallel to the eigenvector corresponding to the third
eigenvalue.

\subsection{Deprojection of numerical halos without observational
  noise}

Having chosen a fiducial ``symmetry'' axis and a line-of-sight, we can
produce synthetic maps of X-ray and Sunyaev-Zel'dovich effect
observations. For that purpose, we use the simulated clusters at a
redshift of $z=0.3$ and project them along the line-of-sight. At $z=0.3$, the cluster sample spans a mass range between $8
\times 10^{14}$ and $1.8 \times 10^{15} \, h^{-1} M_\odot$.

\begin{figure}[ht]
\scalebox{0.45}
{
  \begin{picture}(0,0)
    \includegraphics{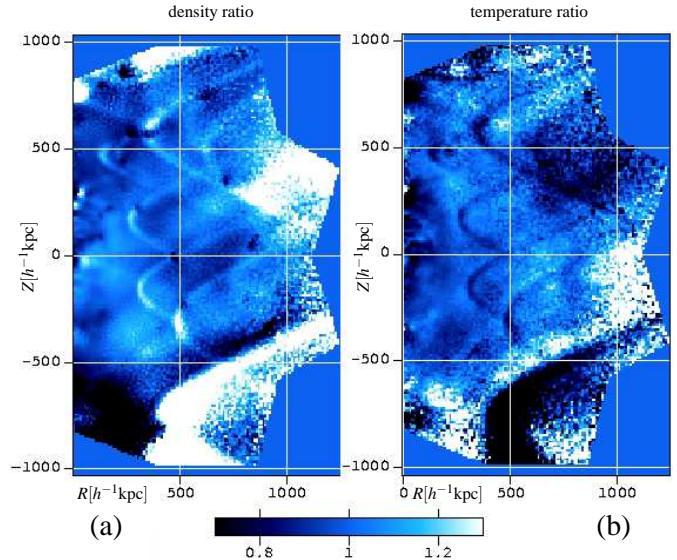}
  \end{picture}
  \setlength{\unitlength}{1mm}
  \begin{picture}(200,170)(0,0)
    \put(58,162){\makebox(0,0){\strut{}\Large density ratio}}
    \put(29,20){\makebox(0,0){\strut{}\Large $R [h^{-1} \textrm{kpc}]$}}
    \put(4,90){\rotatebox{90}{\makebox(0,0){\strut{}\Large $Z [h^{-1} \textrm{kpc}]$}}}
    \put(27,10){\makebox(0,0){\strut{}\Huge (a)}}
    \put(152,162){\makebox(0,0){\strut{}\Large temperature ratio}}
    \put(128,20){\makebox(0,0){\strut{}\Large $R [h^{-1} \textrm{kpc}]$}}
    \put(103,90){\rotatebox{90}{\makebox(0,0){\strut{}\Large $Z [h^{-1} \textrm{kpc}]$}}}
    \put(177,10){\makebox(0,0){\strut{}\Huge (b)}}
  \end{picture}
}
\caption{Reconstruction of the simulated cluster g51 without noise and
  smoothing. The ratios of the reconstructed to the original density
  and temperature are shown. The deprojection was done with $n=5$
  iterations. An inclination angle of $i=68^\circ$ between the
  line-of-sight and the principal inertial axis of the cluster gas was
  chosen and assumed to be known in the reconstruction. Panel (a)
  shows the density ratio $\rho_n/\rho$ and panel (b) the temperature
  ratio $T_n/T$. In the central region, the errors are of the order of
  $10\%$.}
\label{fig:RZratio_g51_noremove}
\end{figure}

\begin{figure}[ht]
\scalebox{0.45}
{
  \begin{picture}(0,0)
    \includegraphics{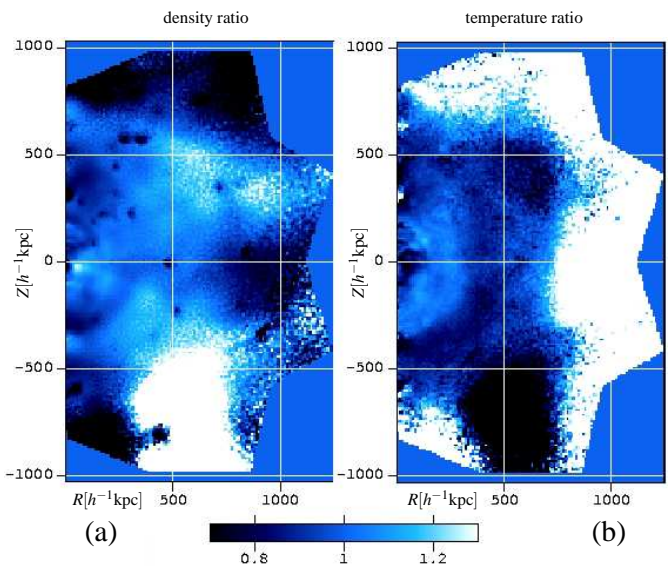}
  \end{picture}
  \setlength{\unitlength}{1mm}
  \begin{picture}(200,170)(0,0)
    \put(58,162){\makebox(0,0){\strut{}\Large density ratio}}
    \put(29,20){\makebox(0,0){\strut{}\Large $R [h^{-1} \textrm{kpc}]$}}
    \put(4,90){\rotatebox{90}{\makebox(0,0){\strut{}\Large $Z [h^{-1} \textrm{kpc}]$}}}
    \put(27,10){\makebox(0,0){\strut{}\Huge (a)}}
    \put(152,162){\makebox(0,0){\strut{}\Large temperature ratio}}
    \put(128,20){\makebox(0,0){\strut{}\Large $R [h^{-1} \textrm{kpc}]$}}
    \put(103,90){\rotatebox{90}{\makebox(0,0){\strut{}\Large $Z [h^{-1} \textrm{kpc}]$}}}
    \put(177,10){\makebox(0,0){\strut{}\Huge (b)}}
  \end{picture}
}
\caption{Reconstruction of the simulated cluster g51 without noise but
  with ``radius-dependent smoothing''. The ratios of the reconstructed
  to the original density and temperature are shown. The deprojection
  was done with $n=5$ iterations. An inclination angle of $i=68^\circ$
  between the line-of-sight and the principal inertial axis of the
  cluster gas was chosen and assumed to be known in the
  reconstruction. Panel (a) shows the density ratio $\rho_n/\rho$ and
  panel (b) the temperature ratio $T_n/T$. In the central region, the
  errors are of the order of $10\%$.}
\label{fig:RZratio_g51}
\end{figure}

For now, we do not add any observational noise to the maps. However,
the clusters contain substructures which break axial symmetry and lead
to artifacts in the density and temperature reconstructions. Thus,
depending on the amount of substructure present in a cluster, it may
still be favourable to use ``radius-dependent smoothing'' on the X-ray and Sunyaev-Zel'dovich effect
maps prior to reconstruction. In Figs.~\ref{fig:RZratio_g51_noremove}
and \ref{fig:RZratio_g51}, we show the results of the deprojection
without any smoothing, and using ``radius-dependent smoothing''.

The density reconstruction in the central region reaches an accuracy
of about $10\%$ in both cases. For the temperature reconstruction and the density reconstruction at large $r$, we obtain somewhat better results without smoothing for this rather symmetric cluster. Note, however,
the hyperbolically shaped artifacts in
Fig.~\ref{fig:RZratio_g51_noremove} which are produced by substructure
clumps in absence of smoothing. They appear at those $R$ and $Z$
values which correspond to the line-of-sight passing through such a
clump. The spike-shaped artifacts discussed in
Sect.~\ref{sec:regularisation} were a special case of the artifacts
found here. For most of the hyperbolae in the left panel of
Fig.~\ref{fig:RZratio_g51_noremove}, one can also see the position of
the clump that produced it in darker colours. The hyperbolae pass
right through them.

As one can see in Fig.~\ref{fig:RZratio_g51}, ``radius-dependent smoothing'' removes the
hyperbolic artifacts. The subclumps, however, still appear in darker
colours in the density ratio map, which means that the reconstructed
density there is too low. However, this is entirely expected and
inevitable, because they violate axial symmetry and thus cannot be
faithfully reconstructed with this deprojection technique. By
smoothing, we essentially remove the subclumps from the maps and
reconstruct the density and temperature of the main halo without them.

Unfortunately, ``radius-dependent smoothing'' also affects the density and temperature
profiles. This can be seen in the ``rings'' around the halo centre in
the right panel of Fig.~\ref{fig:RZratio_g51}. It is further
illustrated in Fig.~\ref{fig:g51_profiles}, which shows the density
and temperature profiles of the original cluster g51, after
deprojection without noise but with ``radius-dependent smoothing'', and after deprojection
without noise and without smoothing. For $r>300 h^{-1} \textrm{kpc}$,
the reconstruction without smoothing yields more accurate density and
temperature profiles. In addition, the profiles for deprojections from
maps including observational noise are shown. They will be discussed
in the next section.
 
Reconstructions along different lines-of-sight and of the three other
clusters in the sample gave similar results. For the most asymmetric
halo, the errors were larger by factors of $1.5$ to $2$ compared to
the reconstruction of g51 presented above. We can thus conclude that,
although clusters are not strictly axisymmetric and contain
substructure, it is possible to apply the deprojection method proposed
in Sect.~\ref{sec:depro_rho_temp} and successfully reconstruct
three-dimensional density and temperature distributions of the cluster
gas.

\subsection{Deprojection of numerical halos including observational
  noise}

In Sect.~\ref{sec:analytic_halos_noise}, we studied the impact of
observational noise in the X-ray and Sunyaev-Zel'dovich effect maps on
the quality of the density and temperature reconstruction. We will now
do the same for the numerically simulated cluster halos using the same
noise model, namely Poisson noise corresponding to $10^4$ observed
photons for the X-ray maps and a noise level expected for a four-hour
ALMA Band 3 observation for the Sunyaev-Zel'dovich effect maps. We
also use the smoothing scheme described there.

\begin{figure}[ht]
\scalebox{0.45}
{
  \begin{picture}(0,0)
    \includegraphics{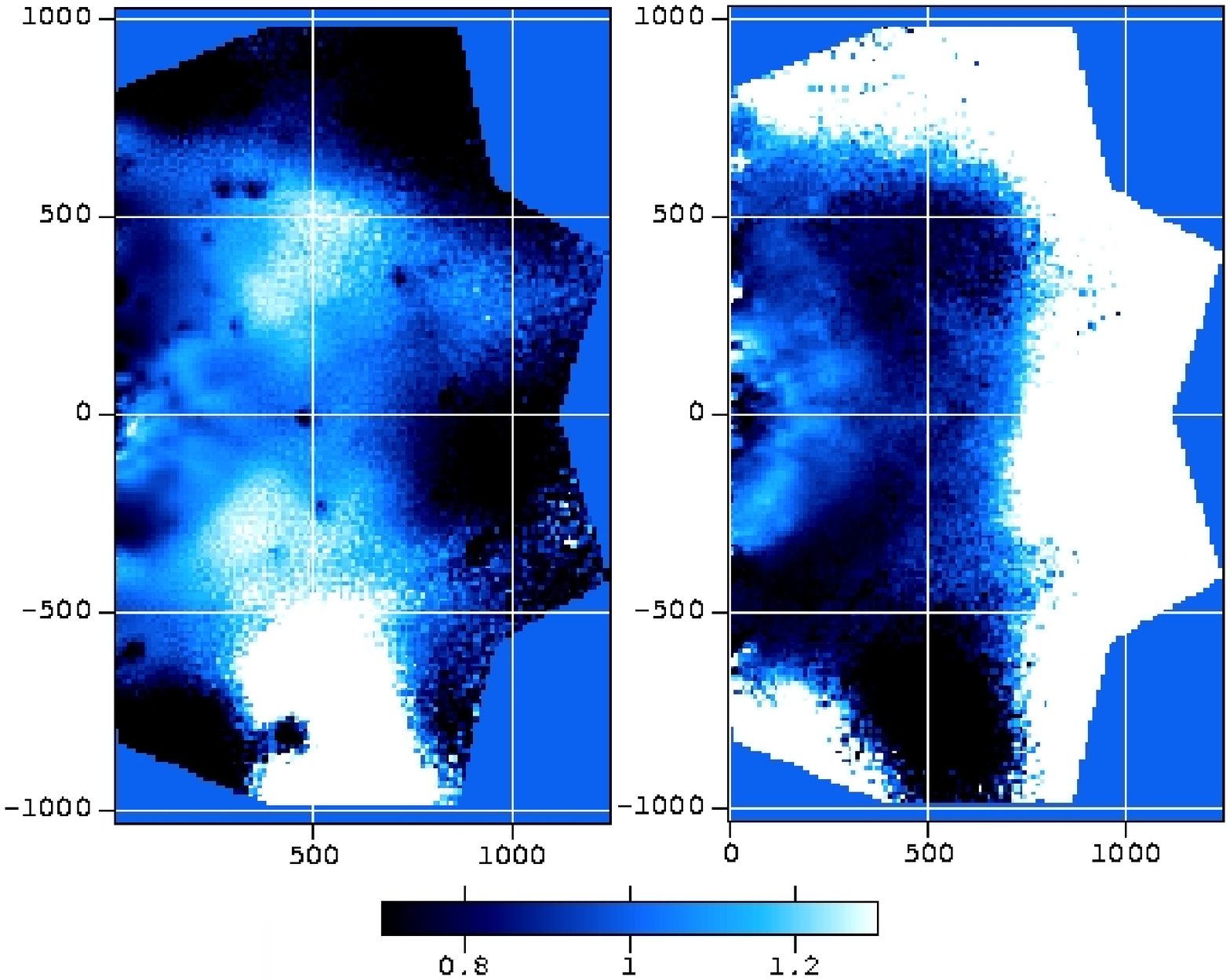}
  \end{picture}
  \setlength{\unitlength}{1mm}
  \begin{picture}(200,170)(0,0)
    \put(58,162){\makebox(0,0){\strut{}\Large density ratio}}
    \put(29,20){\makebox(0,0){\strut{}\Large $R [h^{-1} \textrm{kpc}]$}}
    \put(4,90){\rotatebox{90}{\makebox(0,0){\strut{}\Large $Z [h^{-1} \textrm{kpc}]$}}}
    \put(27,10){\makebox(0,0){\strut{}\Huge (a)}}
    \put(152,162){\makebox(0,0){\strut{}\Large temperature ratio}}
    \put(128,20){\makebox(0,0){\strut{}\Large $R [h^{-1} \textrm{kpc}]$}}
    \put(103,90){\rotatebox{90}{\makebox(0,0){\strut{}\Large $Z [h^{-1} \textrm{kpc}]$}}}
    \put(177,10){\makebox(0,0){\strut{}\Huge (b)}}
  \end{picture}
}
\caption{Reconstruction of the simulated cluster g51 with noise and
  and the complete smoothing scheme applied. The ratios of the reconstructed to the original density
  and temperature are shown. The deprojection was done with $n=5$
  iterations. An inclination angle of $i\approx68^\circ$ between the
  line-of-sight and the principal inertial axis of the cluster gas was
  chosen and assumed to be known in the reconstruction. Panel (a)
  shows the density ratio $\rho_n/\rho$ and panel (b) the temperature
  ratio $T_n/T$. In the central region, the errors are of the order of
  $15\%$.}
\label{fig:RZratio_noise_g51}
\end{figure}

\begin{figure}[ht]
\scalebox{0.7}
{
  \begin{picture}(0,0)%
    \includegraphics{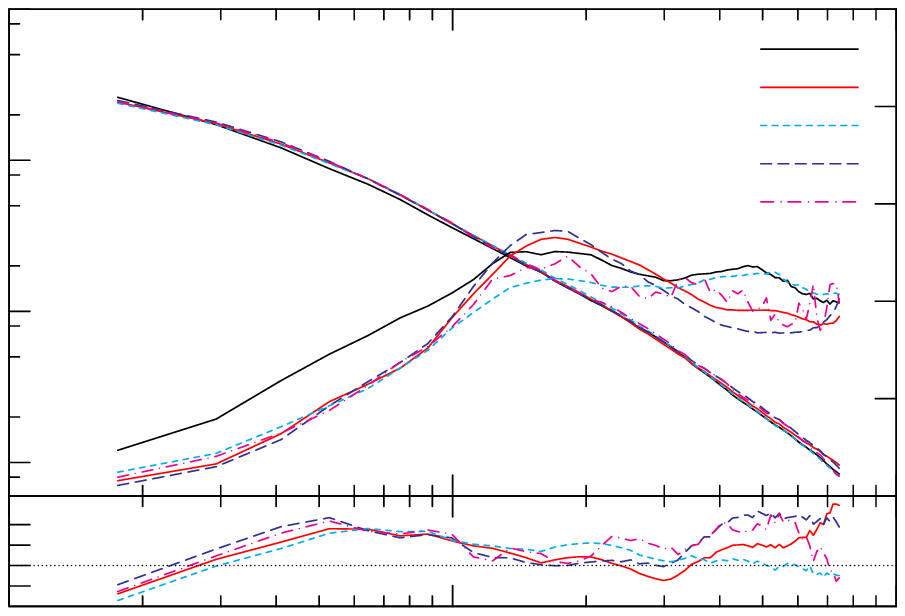}%
  \end{picture}%
    \begingroup
    \setlength{\unitlength}{0.0200bp}%
  \begin{picture}(18000,10800)(0,0)%
    \put(2750,3723){\makebox(0,0)[r]{\strut{} 1e-06}}%
    \put(2750,5898){\makebox(0,0)[r]{\strut{} 1e-05}}%
    \put(2750,8074){\makebox(0,0)[r]{\strut{} 1e-04}}%
    \put(2750,10250){\makebox(0,0)[r]{\strut{} 0.001}}%
    \put(3025,2690){\makebox(0,0){\strut{}}}%
    \put(9413,2690){\makebox(0,0){\strut{}}}%
    \put(15800,2690){\makebox(0,0){\strut{}}}%
    \put(16075,4642){\makebox(0,0)[l]{\strut{} 4}}%
    \put(16075,6044){\makebox(0,0)[l]{\strut{} 6}}%
    \put(16075,7446){\makebox(0,0)[l]{\strut{} 8}}%
    \put(16075,8848){\makebox(0,0)[l]{\strut{} 10}}%
    \put(550,6745){\rotatebox{90}{\makebox(0,0){\strut{} gas density $[10^{10}h^2 M_\odot\textrm{kpc}^{-3}]$}}}%
    \put(17449,6745){\rotatebox{90}{\makebox(0,0){\strut{} gas temperature [keV]}}}%
    \put(13700,9675){\makebox(0,0)[r]{\strut{}original halo}}%
    \put(13700,9125){\makebox(0,0)[r]{\strut{}halo reconstructed without noise}}%
    \put(13700,8575){\makebox(0,0)[r]{\strut{}...and without r.-d. smoothing}}%
    \put(13700,8025){\makebox(0,0)[r]{\strut{}halo reconstructed with noise}}%
    \put(13700,7475){\makebox(0,0)[r]{\strut{}...and without r.-d. sm.}}%
    \put(2750,1944){\makebox(0,0)[r]{\strut{} \scriptsize 0.95}}%
    \put(2750,2239){\makebox(0,0)[r]{\strut{} \scriptsize 1}}%
    \put(2750,2533){\makebox(0,0)[r]{\strut{} \scriptsize 1.05}}%
    \put(2750,2828){\makebox(0,0)[r]{\strut{} \scriptsize 1.1}}%
    \put(3025,1100){\makebox(0,0){\strut{} 10}}%
    \put(9413,1100){\makebox(0,0){\strut{} 100}}%
    \put(15800,1100){\makebox(0,0){\strut{} 1000}}%
    \put(600,2445){\rotatebox{90}{\makebox(0,0){\strut{} density}}}%
    \put(1000,2445){\rotatebox{90}{\makebox(0,0){\strut{} ratio }}}
    \put(1400,2445){\rotatebox{90}{\makebox(0,0){\strut{} $\rho_{n}/\rho$}}}
    \put(9412,275){\makebox(0,0){\strut{} radius $[h^{-1} \textrm{kpc}]$}}%
\end{picture}%
\endgroup
}
\caption{Gas density and temperature profiles of the original and the
  reconstructed cluster g51. The upper panel shows the density
  profiles (falling curves, left axis) and the temperature profiles
  (rising curves, right axis) of the original cluster, the cluster
  reconstructed without observational noise but with ``radius-dependent smoothing'', reconstructed without
  observational noise and without any smoothing,
  reconstructed from maps with observational noise and the complete smoothing scheme applied, and reconstructed
  from maps with observational noise but without ``radius-dependent smoothing''. The lower panel shows the profile of the ratio of the reconstructed density $\rho_n$ to the original density $\rho$. The
  number of iterations used was $n=5$ in all cases. An inclination
  angle of $i\approx68^\circ$ was chosen and assumed to be known in
  the reconstruction.}
\label{fig:g51_profiles}
\end{figure}

We show the results of the reconstruction in
Fig.~\ref{fig:RZratio_noise_g51}. Again, the left panel shows the
ratio of the reconstructed to the original density, and the right
panel the corresponding temperature ratio. In the central region we
achieve an accuracy of about $15\%$. Without ``radius-dependent smoothing'',
errors would be larger by roughly a factor of 1.5 or even more next to
the halo centre. However, if one is mainly interested in density and
temperature profiles it may still be favourable to leave the ``radius-dependent smoothing'' step away. Although the errors are larger without ``radius-dependent smoothing'',
they are less biased with respect to the distance from the halo centre
and cancel better when averaging over spherical shells around it,
especially at large radii. Figure~\ref{fig:g51_profiles} shows the
profiles obtained with and without ``radius-dependent smoothing''.

\begin{figure}[ht]
\scalebox{0.7}
{
  \begin{picture}(0,0)%
    \includegraphics{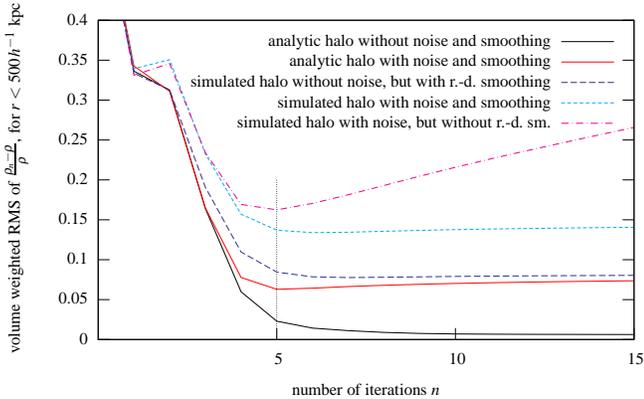}%
  \end{picture}%
    \begingroup
    \setlength{\unitlength}{0.0200bp}%
  \begin{picture}(18000,10800)(0,0)%
    \put(2475,1650){\makebox(0,0)[r]{\strut{} 0}}%
    \put(2475,2725){\makebox(0,0)[r]{\strut{} 0.05}}%
    \put(2475,3800){\makebox(0,0)[r]{\strut{} 0.1}}%
    \put(2475,4875){\makebox(0,0)[r]{\strut{} 0.15}}%
    \put(2475,5950){\makebox(0,0)[r]{\strut{} 0.2}}%
    \put(2475,7025){\makebox(0,0)[r]{\strut{} 0.25}}%
    \put(2475,8100){\makebox(0,0)[r]{\strut{} 0.3}}%
    \put(2475,9175){\makebox(0,0)[r]{\strut{} 0.35}}%
    \put(2475,10250){\makebox(0,0)[r]{\strut{} 0.4}}%
    \put(7558,1100){\makebox(0,0){\strut{} 5}}%
    \put(12367,1100){\makebox(0,0){\strut{} 10}}%
    \put(17175,1100){\makebox(0,0){\strut{} 15}}%
    \put(550,5950){\rotatebox{90}{\makebox(0,0){\strut{}
                   volume weighted RMS of $\frac{\rho_n-\rho}{\rho}$, for $r<500 \, h^{-1}$ kpc}}}%
    \put(9962,275){\makebox(0,0){\strut{}number of iterations $n$}}%
    \put(14950,9675){\makebox(0,0)[r]{\strut{}analytic halo without noise and smoothing}}%
    \put(14950,9125){\makebox(0,0)[r]{\strut{}analytic halo with noise and smoothing}}%
    \put(14950,8575){\makebox(0,0)[r]{\strut{}simulated halo without noise, but with r.-d. smoothing}}%
    \put(14950,8025){\makebox(0,0)[r]{\strut{}simulated halo with noise and smoothing}}%
    \put(14950,7475){\makebox(0,0)[r]{\strut{}simulated halo with noise, but without r.-d. sm.}}%
\end{picture}%
\endgroup
}
\caption{Dependence of the quality of the density reconstruction on
  the number of iterations used. We show the volume-weighted relative
  RMS error $\frac{\rho_n-\rho}{\rho}$ within spheres with $r=500 \,
  h^{-1}$ kpc radius and for different deprojections of our analytic
  halo model and of the numerically simulated cluster g51. The figure legend explains which halo, whether or not noise, and what kind of smoothing were used for the different reconstructions shown. Note that
  ``photon-noise smoothing'' of the X-ray maps is always and only used for maps with noise.}
\label{fig:it_quality}
\end{figure}

We still need to discuss when the iteration used in the density and
temperature reconstructions should best be
stopped. Figure~\ref{fig:it_quality} illustrates the dependence of the
quality of the reconstruction on the number of iterations used. More
precisely, it shows the relative volume-weighted RMS
error of the density reconstruction within $r=500 \, h^{-1}$ kpc as a
function of the number of iterations and for different deprojection
schemes, namely for the deprojections of the analytic halo and the
numerically simulated cluster g51 discussed above and shown in
Figs.~\ref{fig:RZratio_NFW_70}, \ref{fig:RZratio_noise_NFW_70},
\ref{fig:RZratio_g51}, and \ref{fig:RZratio_noise_g51} after $n=20$ or
$n=5$ iterations. The quality of the reconstruction improves quickly
during the first roughly five iterations (first ten for the analytic
halo without noise) and then levels off. In addition, we show the
quality of the reconstruction of g51 from maps with noise but without
using ``radius-dependent smoothing''. In this
case, small-scale noise in the maps is not sufficiently
suppressed. The best reconstruction is found after five
iterations. Then, the quality decreases again because the algorithm
starts to approximate small-scale noise. Thus, unless a halo is very
smooth and axisymmetric, such as the analytic halo without noise, we
find that the quality of the reconstruction does not significantly
increase after $n=5$ iterations and may even decrease if small scale
fluctuations due to noise are not efficiently suppressed. Thus, we
conclude that it is favourable to use this number of iterations for
the deprojection of simulated and real galaxy clusters. Alternatively, one could control the reproduction of small-scale fluctuations with a formal regularisation scheme, such as provided by maximum-entropy methods.

\section{Finding inclination angles}

In all deprojections of analytic and numerical clusters presented
above, we have assumed that the inclination angle of the ``symmetry''
axis is known beforehand. This will usually not be the case for real
observed clusters. In Sect.~\ref{sec:inc_simple}, we sketched how
one can obtain inclination angles by comparing the maps obtained by
projecting reconstructed halos to the original maps. One would
reconstruct a cluster assuming different values for $i$ and find the
value for which the maps match best. In principle, we could generalise
this approach to the X-ray and Sunyaev-Zel'dovich effect maps used for
the density and temperature deprojections. That is, we can compare the
original maps $\psi_{\textrm{X-ray}}$ and $\psi_{\textrm{SZ}}$ to the
maps $\psi_{\textrm{X-ray},N_{\textrm{it}}}$ and
$\psi_{\textrm{SZ},N_{\textrm{it}}}$, which correspond to the
reconstructed halo after a fixed number $N_{\textrm{it}}$ of iterations
and for different values of the inclination angle $i$ used for the
reconstruction.

We did this for the analytic halo model and for our sample of
numerical clusters and used different inclination angles $i'$ for
projecting these halos to obtain the original maps. However, the
minima in the penalty function (see Eq. (\ref{eq:penalty_function})) are not well defined. They are
very broad and not always centred on $i=i'$. Even for the analytic
halo without observational noise, it is hardly possible to find the
correct axis inclination in this way. As one can see from
Eqs.~(\ref{eq:corr_xray}) and (\ref{eq:corr_sz}), the iterative
corrections of the deprojection algorithm are determined from the
deviations of the X-ray and Sunyaev-Zel'dovich effect maps, and the
deviations are thereby minimised. Unfortunately, this still works
remarkably well when choosing a wrong inclination angle $i \neq i'$
for the deprojection. Thus the X-ray and Sunyaev-Zel'dovich effect
maps are still reproduced well in this case, although the errors of
the density and temperature reconstructions increase significantly.

We tried to limit the ability of the deprojection algorithm to
reproduce maps well even when the inclination angle is wrong by
reducing its degrees of freedom. For doing so, we used a variant of
the algorithm that only reconstructs the density and uses a constant
but adjustable temperature. This of course also limits the accuracy of
the reconstruction for the correct inclination angle. Thus, the results
of comparing the reconstructed X-ray and Sunyaev-Zel'dovich effect
maps to the original ones for finding the inclination angle were not
significantly better.

\begin{figure}[ht]
\scalebox{0.7}
{
  \begin{picture}(0,0)%
    \includegraphics{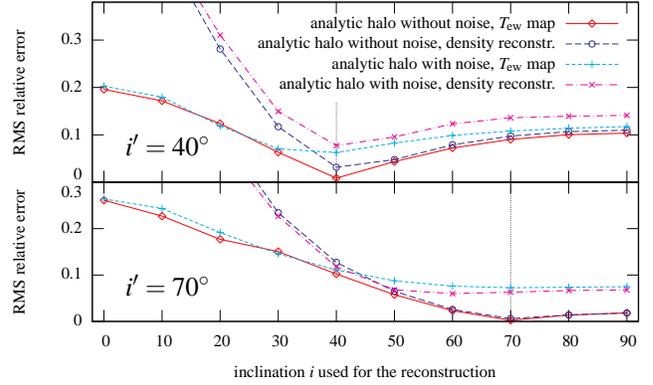}%
  \end{picture}%
    \begingroup
    \setlength{\unitlength}{0.0200bp}%
  \begin{picture}(18000,10800)(0,0)%
    \put(2750,1650){\makebox(0,0)[r]{\strut{} 0}}%
    \put(2750,2900){\makebox(0,0)[r]{\strut{} 0.1}}%
    \put(2750,4150){\makebox(0,0)[r]{\strut{} 0.2}}%
    \put(2750,5100){\makebox(0,0)[r]{\strut{} 0.3}}%
    \put(3338,1100){\makebox(0,0){\strut{} 0}}%
    \put(4902,1100){\makebox(0,0){\strut{} 10}}%
    \put(6465,1100){\makebox(0,0){\strut{} 20}}%
    \put(8029,1100){\makebox(0,0){\strut{} 30}}%
    \put(9593,1100){\makebox(0,0){\strut{} 40}}%
    \put(11157,1100){\makebox(0,0){\strut{} 50}}%
    \put(12721,1100){\makebox(0,0){\strut{} 60}}%
    \put(14285,1100){\makebox(0,0){\strut{} 70}}%
    \put(15848,1100){\makebox(0,0){\strut{} 80}}%
    \put(17412,1100){\makebox(0,0){\strut{} 90}}%
    \put(1100,3525){\rotatebox{90}{\makebox(0,0){\strut{}RMS relative error}}}%
    \put(10375,275){\makebox(0,0){\strut{}inclination $i$ used for the reconstruction}}%
    \put(2750,5600){\makebox(0,0)[r]{\strut{} 0}}%
    \put(2750,6676){\makebox(0,0)[r]{\strut{} 0.1}}%
    \put(2750,7953){\makebox(0,0)[r]{\strut{} 0.2}}%
    \put(2750,9229){\makebox(0,0)[r]{\strut{} 0.3}}%
    \put(3338,4850){\makebox(0,0){\strut{}}}%
    \put(4902,4850){\makebox(0,0){\strut{}}}%
    \put(6465,4850){\makebox(0,0){\strut{}}}%
    \put(8029,4850){\makebox(0,0){\strut{}}}%
    \put(9593,4850){\makebox(0,0){\strut{}}}%
    \put(11157,4850){\makebox(0,0){\strut{}}}%
    \put(12721,4850){\makebox(0,0){\strut{}}}%
    \put(14285,4850){\makebox(0,0){\strut{}}}%
    \put(15848,4850){\makebox(0,0){\strut{}}}%
    \put(17412,4850){\makebox(0,0){\strut{}}}%
    \put(1100,7825){\rotatebox{90}{\makebox(0,0){\strut{}RMS relative error}}}%
    \put(15500,9675){\makebox(0,0)[r]{\strut{}analytic halo without noise, $T_{\textrm{ew}}$ map}}%
    \put(15500,9125){\makebox(0,0)[r]{\strut{}analytic halo without noise, density reconstr.}}%
    \put(15500,8575){\makebox(0,0)[r]{\strut{}analytic halo with noise, $T_{\textrm{ew}}$ map}}%
    \put(15500,8025){\makebox(0,0)[r]{\strut{}analytic halo with noise, density reconstr.}}%
    \put(5000,2550){\makebox(0,0){\strut{} \Large $i'=70^\circ$}}
    \put(5000,6326){\makebox(0,0){\strut{} \Large $i'=40^\circ$}}
\end{picture}%
\endgroup
}
\caption{Accuracy of emission-weighted temperature $T_{\textrm{ew}}$
  maps and densities of reconstructed analytic halos. The
  deprojections started from maps obtained by projecting the analytic
  halo along a line-of-sight with inclination angles of $i'=40^\circ$
  and $i'=70^\circ$. Inclination angles $i$ between $0^\circ$ and
  $90^\circ$ were used for the reconstruction. As expected, the best
  reconstructions are obtained for $i\approx i'$. The errors are shown
  for deprojections from maps without observational noise and from
  smoothed maps with observational noise, and were averaged within a
  region of radius $500 \, h^{-1}$ kpc around the map or halo centre.}
\label{fig:find_inc_analytic}
\end{figure}

\begin{figure}[ht]
\scalebox{0.7}
{
  \begin{picture}(0,0)%
    \includegraphics{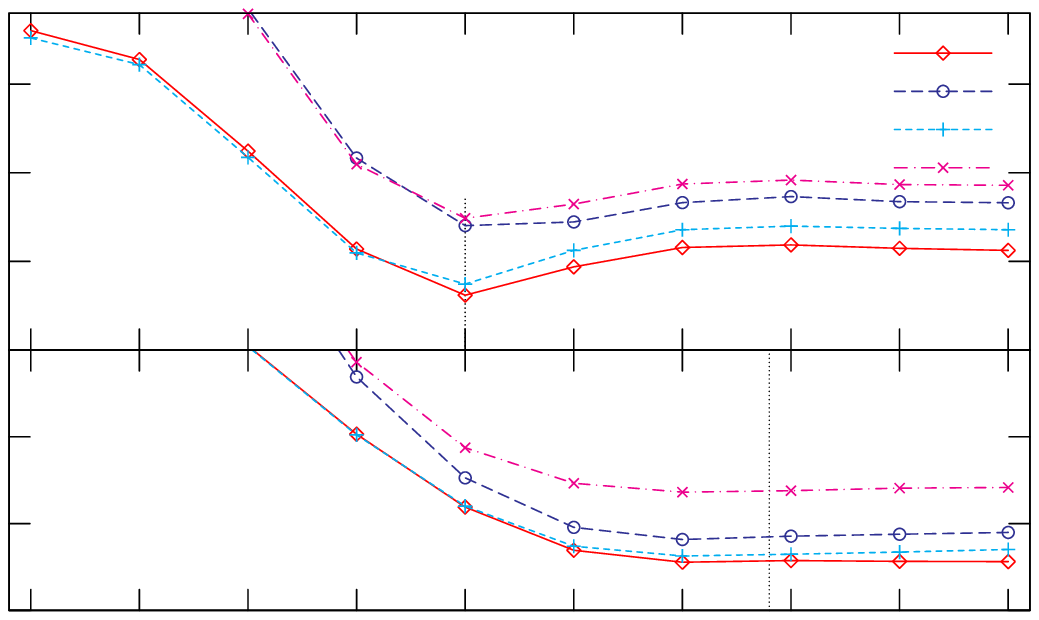}%
  \end{picture}%
    \begingroup
    \setlength{\unitlength}{0.0200bp}%
  \begin{picture}(18000,10800)(0,0)%
    \put(2750,1650){\makebox(0,0)[r]{\strut{} 0}}%
    \put(2750,2900){\makebox(0,0)[r]{\strut{} 0.1}}%
    \put(2750,4150){\makebox(0,0)[r]{\strut{} 0.2}}%
    \put(2750,5100){\makebox(0,0)[r]{\strut{} 0.3}}%
    \put(3338,1100){\makebox(0,0){\strut{} 0}}%
    \put(4902,1100){\makebox(0,0){\strut{} 10}}%
    \put(6465,1100){\makebox(0,0){\strut{} 20}}%
    \put(8029,1100){\makebox(0,0){\strut{} 30}}%
    \put(9593,1100){\makebox(0,0){\strut{} 40}}%
    \put(11157,1100){\makebox(0,0){\strut{} 50}}%
    \put(12721,1100){\makebox(0,0){\strut{} 60}}%
    \put(14285,1100){\makebox(0,0){\strut{} 70}}%
    \put(15848,1100){\makebox(0,0){\strut{} 80}}%
    \put(17412,1100){\makebox(0,0){\strut{} 90}}%
    \put(1100,3525){\rotatebox{90}{\makebox(0,0){\strut{}RMS relative error}}}%
    \put(10375,275){\makebox(0,0){\strut{}inclination $i$ used for the reconstruction}}%
    \put(2750,5600){\makebox(0,0)[r]{\strut{} 0}}%
    \put(2750,6676){\makebox(0,0)[r]{\strut{} 0.1}}%
    \put(2750,7953){\makebox(0,0)[r]{\strut{} 0.2}}%
    \put(2750,9229){\makebox(0,0)[r]{\strut{} 0.3}}%
    \put(3338,4850){\makebox(0,0){\strut{}}}%
    \put(4902,4850){\makebox(0,0){\strut{}}}%
    \put(6465,4850){\makebox(0,0){\strut{}}}%
    \put(8029,4850){\makebox(0,0){\strut{}}}%
    \put(9593,4850){\makebox(0,0){\strut{}}}%
    \put(11157,4850){\makebox(0,0){\strut{}}}%
    \put(12721,4850){\makebox(0,0){\strut{}}}%
    \put(14285,4850){\makebox(0,0){\strut{}}}%
    \put(15848,4850){\makebox(0,0){\strut{}}}%
    \put(17412,4850){\makebox(0,0){\strut{}}}%
    \put(1100,7825){\rotatebox{90}{\makebox(0,0){\strut{}RMS relative error}}}%
    \put(15500,9725){\makebox(0,0)[r]{\strut{}sim. halo without noise, $T_{\textrm{ew}}$ map}}%
    \put(15500,9190){\makebox(0,0)[r]{\strut{}sim. halo without noise, density reconstr.}}%
    \put(15500,8655){\makebox(0,0)[r]{\strut{}sim. halo with noise, $T_{\textrm{ew}}$ map}}%
    \put(15500,8120){\makebox(0,0)[r]{\strut{}sim. halo with noise, density reconstr.}}%
    \put(5000,2550){\makebox(0,0){\strut{} \Large $i'=68^\circ$}}
    \put(5000,6326){\makebox(0,0){\strut{} \Large $i'=40^\circ$}}
\end{picture}%
\endgroup
}
\caption{Accuracy of emission weighted temperature $T_{\textrm{ew}}$
  maps and densities for the reconstructed simulated halo g51. Maps
  obtained by projecting g51 along a line-of-sight with inclination
  angles of $i'=40^\circ$ and $i'=68^\circ$ were used. The
  reconstruction assumed inclination angles $i$ between $0^\circ$ and
  $90^\circ$. Best reconstructions are obtained for $i\approx i'$. The
  errors are shown for deprojections from maps without observational
  noise but using ``radius-dependent smoothing'', and from maps
  with observational noise on which the complete smoothing scheme was
  applied. The RMS relative errors were obtained within a
  circle of radius $200 \, h^{-1}$ kpc around the map centre for the
  $T_{ew}$ maps and inside a sphere of radius $500 \, h^{-1}$ kpc
  around the halo centre for the density reconstructions.}
\label{fig:find_inc_g51}
\end{figure}

On the other hand, leaving the deprojection algorithm as described in
Sect.~\ref{sec:depro_rho_temp}, but using additional independent
information for finding the inclination angle of the halo, seems to be
more promising. For the deprojection, we use maps of the X-ray flux of
the clusters, but so far we do not use any spectral information from
the X-ray observations. In Figs.~\ref{fig:find_inc_analytic} and
\ref{fig:find_inc_g51}, we assume that in addition to the X-ray flux
maps we also have maps of the emission-weighted temperature
$T_{\textrm{ew}}$. We reconstruct the analytic halo and the
numerically simulated cluster g51 from X-ray flux and
Sunyaev-Zel'dovich effect maps as above, but then compare the original
emission-weighted temperature map to one we obtain by projecting the
reconstructed halos. We repeated this for different inclination angles
$i'$, chosen for projecting the original maps, and $i$, chosen in the
reconstruction.

Figure~\ref{fig:find_inc_analytic} shows the RMS relative
error of the reconstructed emission-weighted temperature maps for the
analytic halo without and with noise. $N_{\textrm{it}}=20$ and no
smoothing were used without noise, and $N_{\textrm{it}}=5$ and the
complete smoothing scheme were used with noise. The RMS error was computed
within a radius of $500 \, h^{-1}$ kpc around the map centre and is
shown for inclinations of $i'=40^\circ$ and $i'=70^\circ$ of the
original halo. As desired, the minima of the error curves are at the
correct locations $i\approx i'$.

Note that the curves are only shown for $i$ between $0^\circ$ and
$90^\circ$ because $\psi_{\textrm{X-ray}}$ and $\psi_{\textrm{SZ}}$
and hence the whole deprojection algorithm is insensitive to what is
the front and what is the back side of the cluster. We thus get the
same reconstruction and the same errors for deprojections which adopt
inclination angles $i$ and $180^\circ-i$. There is no way to
distinguish these cases from the X-ray, thermal Sunyaev-Zel'dovich
effect and temperature maps. The error curves are thus symmetric about
$i=90^\circ$. Also note that for the halo with noise, we added
observational noise only to the X-ray flux maps and Sunyaev-Zel'dovich
effect maps that were used for the reconstruction, but not to the
$T_{\textrm{ew}}$ maps which we use for finding the inclination
angle. We do not mimic observational noise in the temperature maps because it can only be realistically modeled when considering instrument response and line emission.
In addition to the error curves of the temperature maps, we
also show in Fig.~\ref{fig:find_inc_analytic} the volume-weighted,
relative RMS errors of the density reconstructions in the central $500
\, h^{-1}$ kpc. As expected, the reconstruction works best for
$i\approx i'$.

Figure~\ref{fig:find_inc_g51} shows similar quantities as
Fig.~\ref{fig:find_inc_analytic}, but for the numerically simulated
cluster halo g51. Original halo inclinations were set to $i'=40^\circ$
and $i'=68^\circ$, and $N_{\textrm{it}}=5$ iterations were used. The
error curves are shown for the simulated halo without observational
noise and using ``radius-dependent smoothing'', and
including observational noise and using the complete smoothing
scheme. No noise was added to the emission-weighted temperature
maps. The relative RMS $T_{\textrm{ew}}$ error was computed within a
circle of radius $200 \, h^{-1}$ kpc around the map centre, while the
density reconstruction errors were again determined within the central
$500 \, h^{-1}$ kpc. Also for this numerical halo the errors are
smallest for $i\approx i'$.

For halos with an original inclination $i'\ll 90^\circ$ or $i'\gg
90^\circ$ and for the analytic halos without noise, the minima of the
error curves for the emission-weighted temperature and the density
reconstructions are well defined. Thus the quality of the
reconstruction of such halos depends strongly on using the correct
inclination $i=i'$ assumed in the deprojection. However, in such cases, the
inclination angle is better constrained by the emission-weighted
temperature maps. On the other hand, if the ``symmetry'' axis of the
original halo is almost perpendicular to the line-of-sight, the minima
of the error curves are usually broad, and finding the precise
inclination $i=i'$ for the reconstruction becomes less important. This
can also be unterstood from the fact that deviations are symmetric around $i=90^\circ$. For example,
the halo with an inclination of $i'=68^\circ$ shown in
Fig.~\ref{fig:find_inc_g51} should exhibit minima at $i=i'=68^\circ$
and at $i=180^\circ-i'=112^\circ$ and a maximum in between. However,
because these three extremal points are close to each other, they
start merging into one broad minimum. Note that the emission-weighted
temperature maps can constrain the inclination angle in both cases to
values where the errors of the reconstruction are close to their
minima.

The accuracy of inclination-angle estimates could be improved by using
other independent information in addition to the temperature maps,
such as data from weak and strong-lensing observations. Lensing
observations allow reconstructions of the lensing potential \citep[see
e.g.][]{CA05.1}, which is simply the suitably rescaled projection of
the lens' gravitational potential. A natural way to employ this for
finding inclination angles is to assume hydrostatic equilibrium of the
cluster gas in the potential well of the cluster and use the density
and temperature reconstruction described above to obtain the
gravitational potential of the reconstructed cluster halo. Its
projection can then be compared to the lensing potential obtained from
observations. Alternatively, one could use the deprojection algorithm
discussed in Sect.~\ref{sec:depro_quantity} to obtain the
three-dimensional gravitational potential from the lensing potential
that was found from observations and compare it to the gravitational
potential corresponding to the density and temperature reconstruction
under the assumption of hydrostatic equilibrium. We shall
explore these possibilities in a forthcoming study.

\section{Summary and discussion}

We propose a new method for deprojecting hot gas in galaxy clusters
which combines X-ray and thermal Sunyaev-Zel'dovich effect
observations and reconstructs three-dimensional density and
temperature distributions. We start from the iterative deprojection
algorithm suggested by \cite{BI90.1} which employs Richardson-Lucy
deconvolution and assumes axial symmetry of the physical quantity
whose three-dimensional distribution shall be reconstructed from
two-dimensional maps of its projection along the line-of-sight.

This approach does not restrict the orientation of the symmetry axis
to be parallel to the line-of-sight, but the inclination angle between
the symmetry axis and the line-of-sight is assumed to be known. This
algorithm runs into problems when it is used to reconstruct strongly
peaked distributions such as the X-ray emissivity of a galaxy
cluster. There, one obtains spike-shaped artifacts through the centre
of the reconstructed halo. We suppress the formation of such artifacts
by introducing a regularisation scheme for the iterative corrections
used in the deprojection. Then, we generalise this algorithm to
simultaneously reconstruct three-dimensional distributions of several
physical quantities by combining two-dimensional maps of projections
which probe these three-dimensional distributions in different ways.

Here, we discuss how three-dimensional density and temperature
distributions of the intra-cluster medium can be reconstructed from
combined X-ray and thermal Sunyaev-Zel'dovich effect observations. We
test the method using synthetic data of analytically modeled and of
numerically simulated galaxy clusters and discuss the quality of the
reconstructions and the impact of observational noise, cluster
substructure and deviations from axial symmetry. For numerical
clusters which are of course not strictly axisymmetric, we use one of
the principal inertial axes as the ``symmetry'' axis for the
deprojection.

Our main findings, if we neglect observational noise and assume that
the inclination angle between the symmetry axis and the line-of-sight
is known, are:

\begin{itemize}

\item Spike-shaped artifacts of the deprojection are efficiently
  suppressed by the regularisation of the iterative corrections.

\item Densities and temperatures of the ICM of axisymmetric analytic
  clusters can be reconstructed very accurately from X-ray flux and
  Sunyaev-Zel'dovich effect maps. Errors are of the order of 1\%
  unless the angle between the symmetry axis and the line-of-sight is
  small.

\item The three-dimensional density and temperature distributions of
  hot gas in numerically simulated clusters, although not strictly
  axisymmetric, can still be reliably reconstructed. Relative errors
  reach roughly 10\%. Smoothing of the X-ray flux and the
  Sunyaev-Zel'dovich effect maps can be used to suppress artifacts
  caused by subclumps.

\item Accurate gas density and temperature profiles can be obtained
  from the reconstructions.

\end{itemize}

We then add photon noise corresponding to a total number of $10^4$
observed photons to the X-ray and observational noise corresponding to
a four-hour ALMA Band 3 observation to the Sunyaev-Zel'dovich effect
maps, respectively. We smooth the maps before repeating the
deprojections to suppress small-scale fluctuations which the algorithm
would otherwise attempt to approximate and thereby reduce the quality of the
reconstructions. From the repeated reconstructions of the analytic and
numerical halos, this time including observational noise, we conclude:

\begin{itemize}

\item Gas densities and temperatures of axisymmetric analytic halos
  can also be efficiently reconstructed from maps that contain
  observational noise. The relative errors of the reconstructions are
  about 5\% to 10\%.

\item The three-dimensional structure of the ICM of numerically
  simulated clusters can also be reliably reconstructed from X-ray
  flux and Sunyaev-Zel'dovich effect maps that contain observational
  noise. Relative errors reach roughly 15\%.
  
\item Accurate profiles can be obtained from the reconstructions. 
  
\item Five iterations are sufficient for the ICM
  deprojection. Using a larger number does not increase the quality of the reconstruction.       
    
\end{itemize}  

For these deprojections, we assumed that the inclination angle between
the symmetry axis and the line-of-sight is known beforehand. This will
usually not be the case for observations of real clusters. In
principle, one could try to find the inclination angle by deprojecting
the cluster assuming different values of the inclination angle and
comparing the original X-ray and Sunyaev-Zel'dovich effect maps to
those expected from the reconstructed halo. They should match best if
the correct inclination is used for the deprojection. Unfortunately,
the minima of the deviations of the maps as a function of the
assumed inclination angle are quite broad and not always centred on
the correct value. However additional data which are independent of
the X-ray flux and the Sunyaev-Zel'dovich effect maps, can provide
information on the inclination angle. We show that high-quality
emission-weighted temperature maps which become more and more
routinely available, can constrain the inclination angle of a
cluster's symmetry axis to values for which the quality of the
reconstruction is close to its optimum.

\acknowledgements{We are deeply indebted to Klaus Dolag, who generously provided us access to the numerical simulations of the cluster sample that was used in this work. We also thank Massimo Meneghetti for useful discussions. E.~P.~is supported by the German Science Foundation under grant number BA~1369/6-1.}

\bibliographystyle{aa}
\bibliography{master}

\end{document}